\documentclass[twoside,11pt]{article}

\usepackage{jmlr2e}
\usepackage{amsmath}
\usepackage{algorithm}
\usepackage{algorithmic}
\usepackage{subfigure}
\usepackage{graphicx}
\usepackage{bbm}
\usepackage{tikz}
\usetikzlibrary{arrows}
\usetikzlibrary{graphs}
\usepackage{url}
\usepackage{varwidth}
\usepackage{multirow}

\ShortHeadings{IFM LAB TUTORIAL SERIES \# 5, COPYRIGHT \copyright IFM LAB}{JIAWEI ZHANG, IFM LAB DIRECTOR}
\firstpageno{1}

\begin{document}

\title{Basic Neural Units of the Brain: Neurons, Synapses and Action Potential}

\author{\name Jiawei Zhang \email jiawei@ifmlab.org \\
	\addr{Founder and Director}\\
       {Information Fusion and Mining Laboratory}\\
       (First Version: May 2019; Revision: May 2019.)}

\maketitle

\begin{abstract}

As a follow-up tutorial article of \cite{zhang2019secrets}, in this paper, we will introduce the basic compositional units of the human brain, which will further illustrate the cell-level bio-structure of the brain. On average, the human brain contains about 100 billion neurons and many more neuroglia which serve to support and protect the neurons. Each neuron may be connected to up to 10,000 other neurons, passing signals to each other via as many as 1,000 trillion synapses. In the nervous system, a synapse is a structure that permits a neuron to pass an electrical or chemical signal to another neuron or to the target effector cell. Such signals will be accumulated as the membrane potential of the neurons, and it will trigger and pass the signal pulse (i.e., action potential) to other neurons when the membrane potential is greater than a precisely defined threshold voltage. To be more specific, in this paper, we will talk about the neurons, synapses and the action potential concepts in detail. Many of the materials used in this paper are from wikipedia and several other neuroscience introductory articles, which will be properly cited in this paper. This is the second of the three tutorial articles about the brain (the other two are \cite{zhang2019secrets} and \cite{zhang2019cognitive}). The readers are suggested to read the previous tutorial article \cite{zhang2019secrets} to get more background information about the brain structure and functions prior to reading this paper.

\end{abstract}


\tableofcontents

\section{Introduction}

According to \cite{neuron_synapse}, the core component of the nervous system in general, and the brain in particular, is the neuron. A neuron is an electrically excitable cell that processes and transmits information by electro-chemical signalling. Unlike other cells, neurons never divide, and neither do they die off to be replaced by new ones. By the same token, they usually cannot be replaced after being lost, although there are a few exceptions. The average human brain has about 100 billion neurons (or nerve cells) and many more neuroglia which serve to support and protect the neurons. Each neuron may be connected to up to 10,000 other neurons, passing signals to each other via as many as 1,000 trillion synaptic connections, equivalent by some estimates to a computer with a 1 trillion bit per second processor. Estimates of the human brain's memory capacity vary wildly from 1 to 1,000 terabytes (for comparison, the 19 million volumes in the US Library of Congress represents about 10 terabytes of data). Information transmission within the brain, such as takes place during the processes of memory encoding and retrieval, is achieved using a combination of chemicals and electricity.

\begin{figure}[t]
    \centering
    \includegraphics[width=0.8\textwidth]{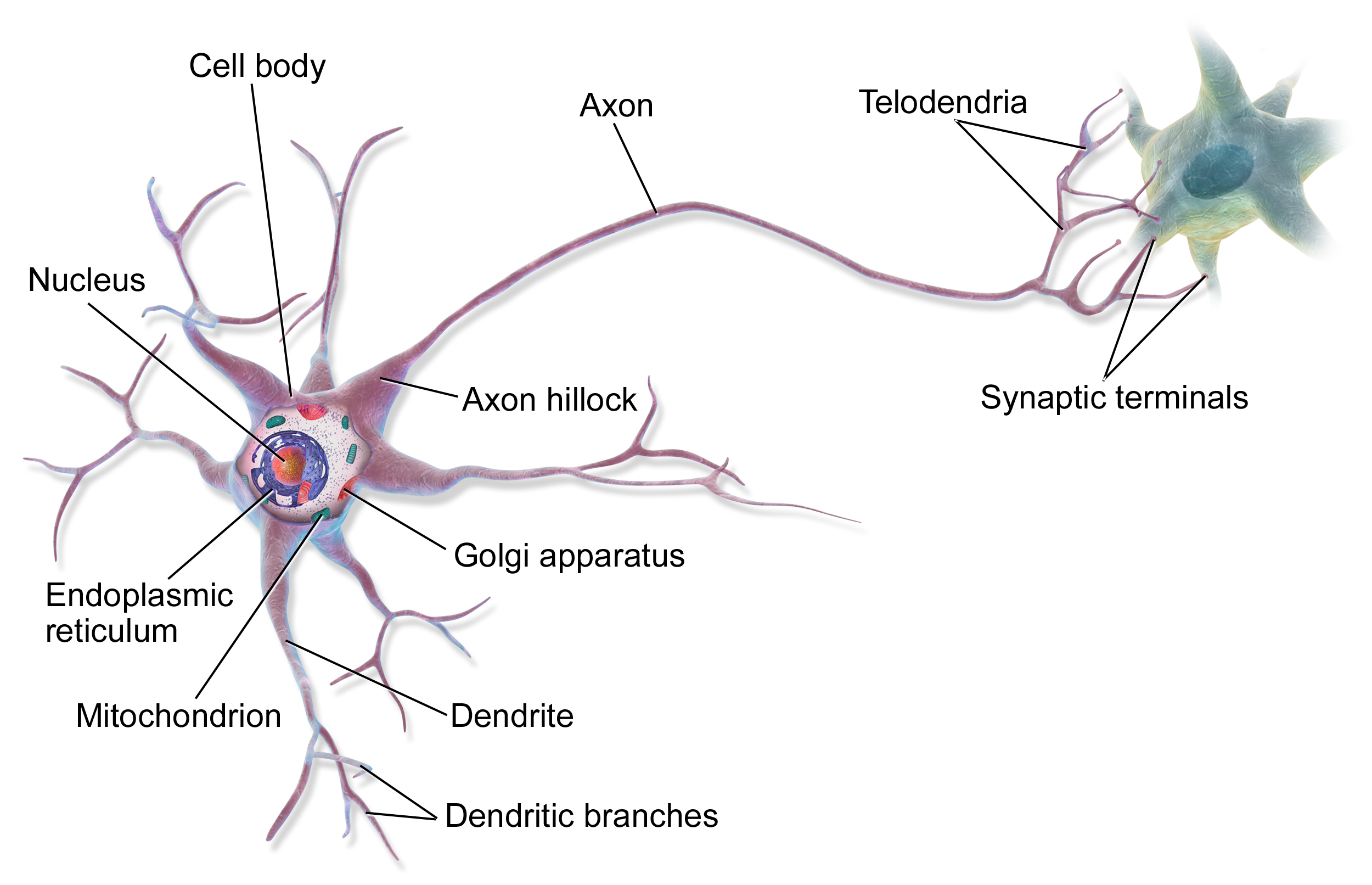}
    \caption{An Illustration of the Brain Neuron \cite{neuron}.}
    \label{fig:brain_cell_neuron}
\end{figure}

As illustrated in Figure~\ref{fig:brain_cell_neuron}, a typical neuron possesses a soma (the bulbous cell body which contains the cell nucleus), dendrites (long, feathery filaments attached to the cell body in a complex branching ``dendritic tree'') and a single axon (a special, extra-long, branched cellular filament, which may be thousands of times the length of the soma). Every neuron maintains a voltage gradient across its membrane, due to metabolically-driven differences in ions of sodium, potassium, chloride and calcium within the cell, each of which has a different charge. If the voltage changes significantly, an electrochemical pulse called an action potential (or nerve impulse) is generated. This electrical activity can be measured and displayed as a wave form called brain wave or brain rhythm.

This pulse travels rapidly along the cell's axon, and is transferred across a specialized connection known as a synapse to a neighboring neuron, which receives it through its feathery dendrites. A synapse is a complex membrane junction or gap (the actual gap, also known as the synaptic cleft, is of the order of 20 nanometres, or 20 millionths of a millimetre) used to transmit signals between cells, and this transfer is therefore known as a synaptic connection. Although axon-dendrite synaptic connections are the norm, other variations (e.g. dendrite-dendrite, axon-axon, dendrite-axon) are also possible. A typical neuron fires 5 - 50 times every second.

Each individual neuron can form thousands of links with other neurons in this way, giving a typical brain well over 100 trillion synapses (up to 1,000 trillion, by some estimates). Functionally related neurons connect to each other to form neural networks (also known as neural nets or assemblies). The connections between neurons are not static, though, they change over time. The more signals sent between two neurons, the stronger the connection grows (technically, the amplitude of the post-synaptic neuron's response increases), and so, with each new experience and each remembered event or fact, the brain slightly re-wires its physical structure.

\begin{figure}[t]
    \centering
    \includegraphics[width=0.8\textwidth]{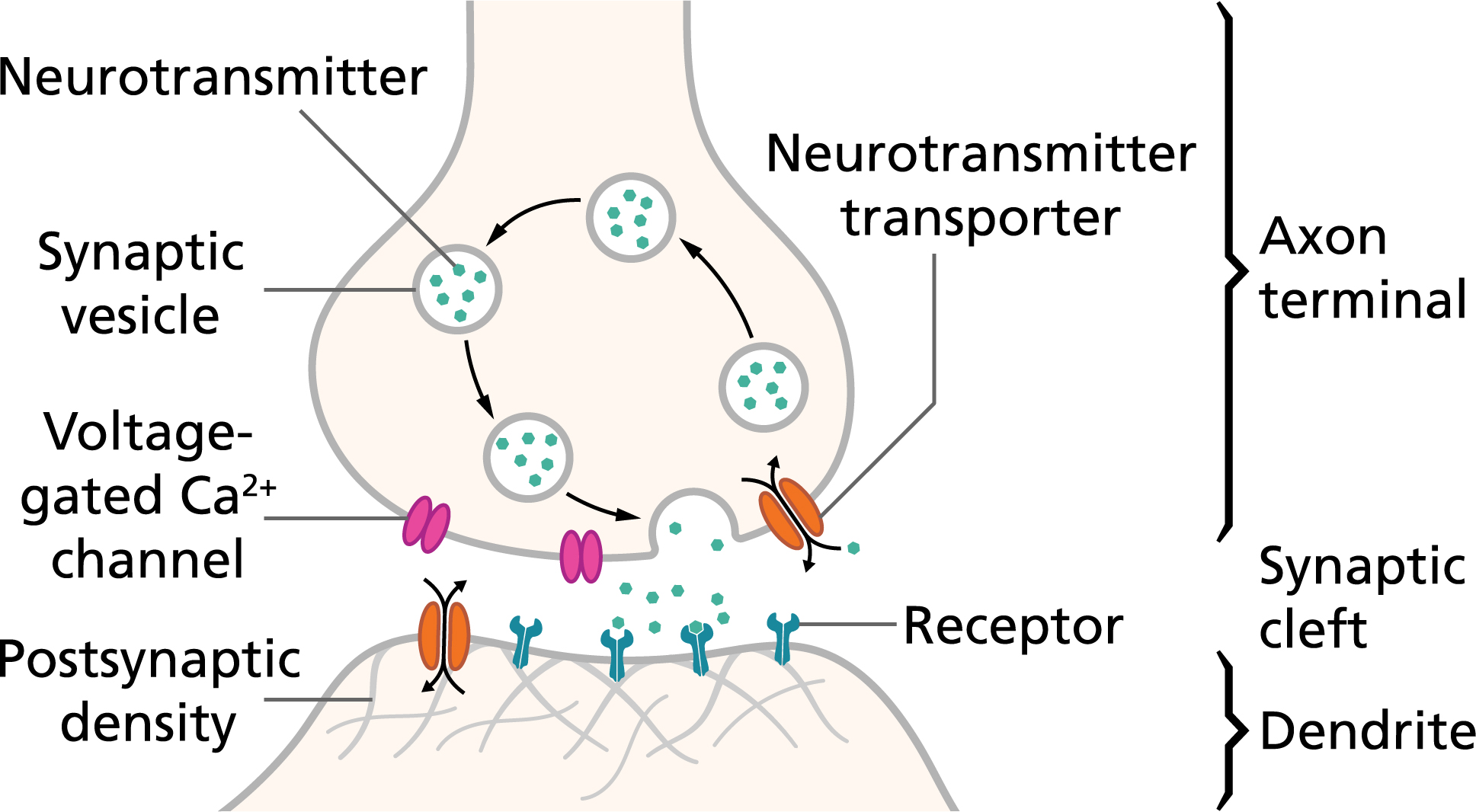}
    \caption{An Illustration of the Brain Synapses \cite{synapse}.}
    \label{fig:brain_cell_synapse}
\end{figure}

The interactions of neurons is not merely electrical, though, but electro-chemical. As illustrated in Figure~\ref{fig:brain_cell_synapse}, each axon terminal contains thousands of membrane-bound sacs called vesicles, which in turn contain thousands of neurotransmitter molecules each. Neurotransmitters are chemical messengers which relay, amplify and modulate signals between neurons and other cells. The two most common neurotransmitters in the brain are the \textit{amino acids glutamate} and GABA (i.e., \textit{gamma-Aminobutyric acid}, or \textit{$\gamma$-aminobutyric acid}); other important neurotransmitters include \textit{acetylcholine}, \textit{dopamine}, \textit{adrenaline}, \textit{histamine}, \textit{serotonin} and \textit{melatonin}.

When stimulated by an electrical pulse, neurotransmitters of various types are released, and they cross the cell membrane into the synaptic gap between neurons. These chemicals then bind to chemical receptors in the dendrites of the receiving (post-synaptic) neuron. In the process, they cause changes in the permeability of the cell membrane to specific ions, opening up special gates or channels which let in a flood of charged particles (ions of calcium, sodium, potassium and chloride). This affects the potential charge of the receiving neuron, which then starts up a new electrical signal in the receiving neuron. The whole process takes less than one five-hundredth of a second. In this way, a message within the brain is converted, as it moves from one neuron to another, from an electrical signal to a chemical signal and back again, in an ongoing chain of events which is the basis of all brain activity.

The electro-chemical signal released by a particular neurotransmitter may be such as to encourage to the receiving cell to also fire, or to inhibit or prevent it from firing. Different neurotransmitters tend to act as excitatory (e.g. acetylcholine, glutamate, aspartate, noradrenaline, histamine) or inhibitory (e.g. GABA, glycine, seratonin), while some (e.g. dopamine) may be either. 

As has been mentioned, in addition to neurons, the brain contains about an equal mass of neuroglia, the most common types being oligodendrocytes, astrocytes and microglia. Because they are so much smaller than neurons, there are up to 10 times as many in number, and different areas of the brain have higher or lower concentrations of glia. It used to be thought that the role of neuroglia was limited to the physical support, nutrition and repair of the neurons of the central nervous system. However, more recent research suggests that glia, particularly astrocytes, actually perform a much more active role in brain communication and neuroplasticity, although the extent and mechanics of of this role is still uncertain, and a substantial amount of contemporary brain research is now focused on glials cells.

The above descriptions provide a brief introduction to the cell-level brain working mechanism and process. To help the readers understand such a process, in the following parts of this paper, we will focus on introducing the \textit{neurons}, \textit{synapses}, \textit{neuroglia} and \textit{action potential} specifically. Many of the terms mentioned above will be explained in great detail as follows.

\section{Neuron}

According to \cite{neuron}, Neurons are the primary components of the nervous system, along with the neuroglia that give them structural and metabolic support. The nervous system is made up of the central nervous system, which includes the brain and spinal cord, and the peripheral nervous system, which includes the autonomic and somatic nervous systems. In vertebrates, the majority of neurons belong to the central nervous system, but some reside in peripheral ganglia, and many sensory neurons are situated in sensory organs such as the retina and cochlea. All animals except sponges and placozoans have neurons, but other multicellular organisms such as plants do not.

A typical neuron consists of a cell body (soma), dendrites, and a single axon. The soma is usually compact. The axon and dendrites are filaments that extrude from it. Dendrites typically branch profusely and extend a few hundred micrometers from the soma. The axon leaves the soma at a swelling called the axon hillock, and travels for as far as 1 meter in humans or more in other species. It branches but usually maintains a constant diameter. At the farthest tip of the axon's branches are axon terminals, where the neuron can transmit a signal across the synapse to another cell. Neurons may lack dendrites or have no axon. Most neurons receive signals via the dendrites and soma and send out signals down the axon. At the majority of synapses, signals cross from the axon of one neuron to a dendrite of another. However, synapses can connect an axon to another axon or a dendrite to another dendrite.

Neurons are electrically excitable and can communicate with other cells via specialized connections called synapses. It is the main component of nervous tissue. The signaling process is partly electrical and partly chemical. Neurons maintain the voltage gradients across their membranes. If the voltage changes by a large enough amount over a short interval, the neuron generates an all-or-nothing electrochemical pulse called an action potential. This potential travels rapidly along the axon, and activates synaptic connections as it reaches them. Synaptic signals may be excitatory or inhibitory, increasing or reducing the net voltage that reaches the soma.

In this section, we will talk about the neuron, and we will introduce its cell structure, its structural classification, functional classification, the connectivity and its coding scheme, respectively.

\subsection{Neuron Cell Structure}

\begin{figure}[t]
    \centering
    \includegraphics[width=0.8\textwidth]{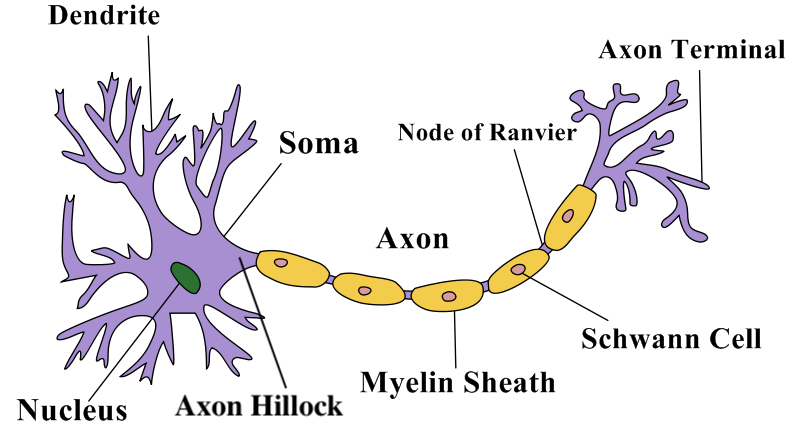}
    \caption{An Illustration of the Brain Neuron Cell Structure \cite{neuron}.}
    \label{fig:brain_cell_neuron_2}
\end{figure}

\subsubsection{Neuron Cell Structure} 

As introduced in \cite{neuron}, neurons are highly specialized for the processing and transmission of cellular signals. Given their diversity of functions performed in different parts of the nervous system, there is a wide variety in their shape, size, and electrochemical properties. For instance, the soma of a neuron can vary from 4 to 100 micrometers in diameter. As illustrated in Figure~\ref{fig:brain_cell_neuron_2}, the neuron cell body includes several important structures, including the \textit{soma}, \textit{nucleus}, \textit{dendrite}, \textit{axon}, \textit{myelin sheath}, \textit{Schwann cell}, \textit{node of Ranvier} and \textit{axon terminal}.

\begin{itemize}
\item \textbf{Soma}: The soma is the body of the neuron. As it contains the nucleus, most protein synthesis occurs here. The nucleus can range from 3 to 18 micrometers in diameter.
\item \textbf{Dendrites}: The dendrites of a neuron are cellular extensions with many branches. This overall shape and structure is referred to metaphorically as a dendritic tree. This is where the majority of input to the neuron occurs via the dendritic spine.
\item \textbf{Axon}: The axon is a finer, cable-like projection that can extend tens, hundreds, or even tens of thousands of times the diameter of the soma in length. The axon primarily carries nerve signals away from the soma, and carries some types of information back to it. Many neurons have only one axon, but this axon may - and usually will - undergo extensive branching, enabling communication with many target cells. 
\item \textbf{Axon Hillock}: The part of the axon where it emerges from the soma is called the axon hillock. Besides being an anatomical structure, the axon hillock also has the greatest density of voltage-dependent sodium channels. This makes it the most easily excited part of the neuron and the spike initiation zone for the axon. In electrophysiological terms, it has the most negative threshold potential. While the axon and axon hillock are generally involved in information outflow, this region can also receive input from other neurons.
\item \textbf{Myelin}: Myelin is a lipid-rich substance formed in the central nervous system by neuroglia called \textit{oligodendrocytes}, and in the peripheral nervous system by Schwann cells. Myelin insulates nerve cell axons to increase the speed at which information travels from one nerve cell body to another or, for example, from a nerve cell body to a muscle. The myelinated axon can be likened to an electrical wire with insulating material (myelin) around it.
\item \textbf{Node of Ranvier}: Unlike the plastic covering on an electrical wire, myelin does not form a single long sheath over the entire length of the axon. Rather, each myelin sheath insulates the axon over a single section and, in general, each axon comprises multiple long myelinated sections separated from each other by short gaps called Nodes of Ranvier.
\item \textbf{Axon Terminal}: The axon terminal is found at the end of the axon farthest from the soma and contains synapses. Synaptic boutons are specialized structures where neurotransmitter chemicals are released to communicate with target neurons.
\end{itemize}

The accepted view of the neuron attributes dedicated functions to its various anatomical components; however, dendrites and axons often act in ways contrary to their so-called main function. Diagram of a typical myelinated vertebrate motor neuron axons and dendrites in the central nervous system are typically only about one micrometer thick, while some in the peripheral nervous system are much thicker. The soma is usually about 10-25 micrometers in diameter and often is not much larger than the cell nucleus it contains. The longest axon of a human motor neuron can be over a meter long, reaching from the base of the spine to the toes.

For the real-world neurons, the axons and dendrites can be very similar in structure, which makes it difficult to distinguish them form each other. Based on the above descriptions, we also would like to add a remark about the differences between the axon and the dendrite, which are serve as the guidance to differentiate axons from dendrites.

\hspace{-2em}
\begin{varwidth}[t]{.55\textwidth}
\textbf{Axons}
\begin{itemize}
\item Take information away from the cell body
\item Smooth Surface
\item Generally only 1 axon per cell
\item No ribosomes
\item Can have myelin
\item Branch further from the cell body
\end{itemize}
\end{varwidth}
\hspace{2em}
\begin{varwidth}[t]{.5\textwidth}
\textbf{Dendrites}
\begin{itemize}
\item Bring information to the cell body
\item Rough Surface (dendritic spines)
\item Usually many dendrites per cell
\item Have ribosomes
\item No myelin insulation
\item Branch near the cell body
\end{itemize}
\end{varwidth}

\subsubsection{Histology and Internal Structure}

\begin{figure}[t]
    \centering
    \includegraphics[width=0.8\textwidth]{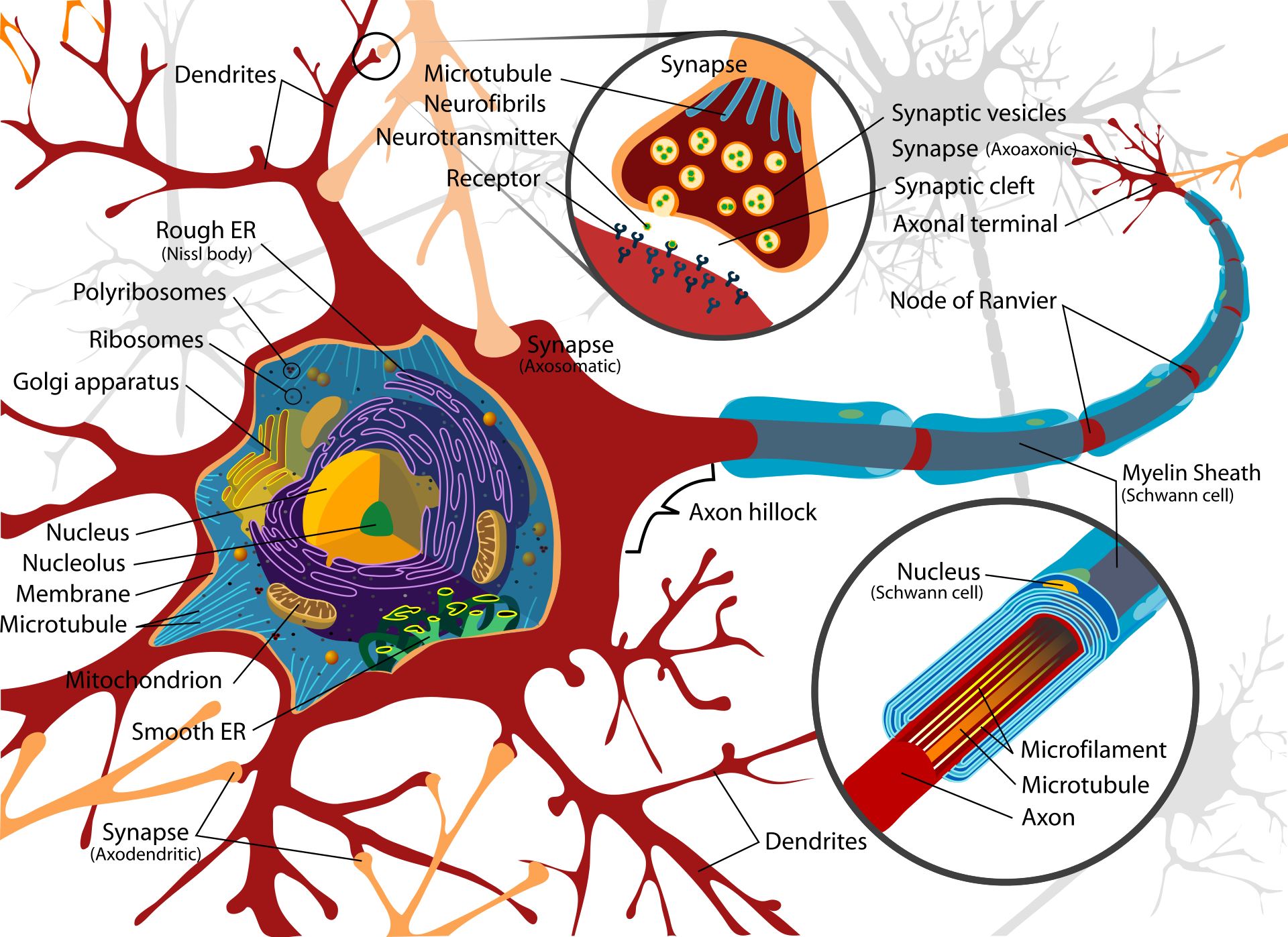}
    \caption{An Illustration of the Inner Structure of Neuron Cell \cite{neuron}.}
    \label{fig:brain_cell_neuron_inner_structure}
\end{figure}

Besides these basic structures, the neurons can contain many other structures inside their cell body (i.e., soma). As introduced in \cite{neuron}, numerous microscopic clumps called Nissl bodies (or Nissl substance, whose structure and position is illustrated in Figure~\ref{fig:brain_cell_neuron_inner_structure}) are seen when nerve cell bodies are stained with a basophilic dye. These structures consist of rough endoplasmic reticulum and associated ribosomal RNA. Named after German psychiatrist and neuropathologist Franz Nissl, they are involved in protein synthesis and their prominence can be explained by the fact that nerve cells are very metabolically active. Basophilic dyes such as aniline or (weakly) haematoxylin highlight negatively charged components, and so bind to the phosphate backbone of the ribosomal RNA.

The cell body of a neuron is supported by a complex mesh of structural proteins called neurofilaments, which together with neurotubules (neuronal microtubules) are assembled into larger neurofibrils as illustrated at the right bottom of Figure~\ref{fig:brain_cell_neuron_inner_structure}. Some neurons also contain pigment granules, such as neuromelanin (a brownish-black pigment that is byproduct of synthesis of catecholamines), and lipofuscin (a yellowish-brown pigment), both of which accumulate with age. Other structural proteins that are important for neuronal function are actin and the tubulin of microtubules. Actin is predominately found at the tips of axons and dendrites during neuronal development. There the actin dynamics can be modulated via an interplay with microtubule.

There are different internal structural characteristics between axons and dendrites. Typical axons almost never contain ribosomes, except some in the initial segment. Dendrites contain granular endoplasmic reticulum or ribosomes, in diminishing amounts as the distance from the cell body increases.

\subsubsection{Membrane}\label{subsec:neuron_membrane}

\begin{figure}[t]
    \centering
    \includegraphics[width=0.8\textwidth]{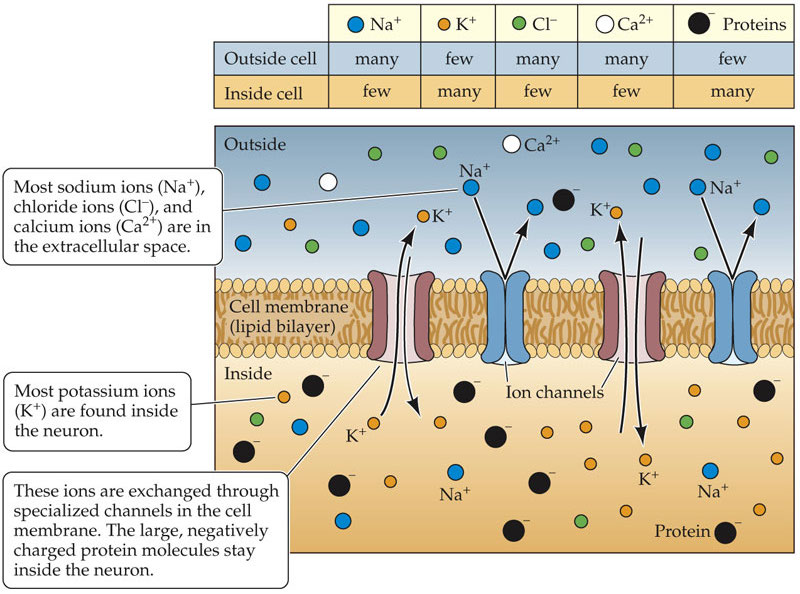}
    \caption{An Illustration of the Neuron Membrane \cite{membrane}.}
    \label{fig:brain_cell_membrane}
\end{figure}

Like all animal cells, the cell body of every neuron is enclosed by a plasma membrane, a bilayer of lipid molecules with many types of protein structures embedded in it. A lipid bilayer is a powerful electrical insulator, but in neurons, many of the protein structures embedded in the membrane are electrically active. As illustrated in Figure~\ref{fig:brain_cell_membrane}, these include ion channels that permit electrically charged ions to flow across the membrane and ion pumps that chemically transport ions from one side of the membrane to the other. 

Most ion channels are permeable only to specific types of ions. Some ion channels are voltage gated, meaning that they can be switched between open and closed states by altering the voltage difference across the membrane. Others are chemically gated, meaning that they can be switched between open and closed states by interactions with chemicals that diffuse through the extracellular fluid. The ion materials include sodium, potassium, chloride, and calcium.The interactions between ion channels and ion pumps produce a voltage difference across the membrane, typically a bit less than 1/10 of a volt at baseline. This voltage has two functions: first, it provides a power source for an assortment of voltage-dependent protein machinery that is embedded in the membrane; second, it provides a basis for electrical signal transmission between different parts of the membrane.

We will introduce more about the neuron membrane and membrane potential, as well as the working mechanism of neurons to maintain the membrane potential in Section~\ref{sec:action_potential}.

\subsection{Neuron Structural Classification}

\begin{figure}[t]
    \centering
    \includegraphics[width=0.8\textwidth]{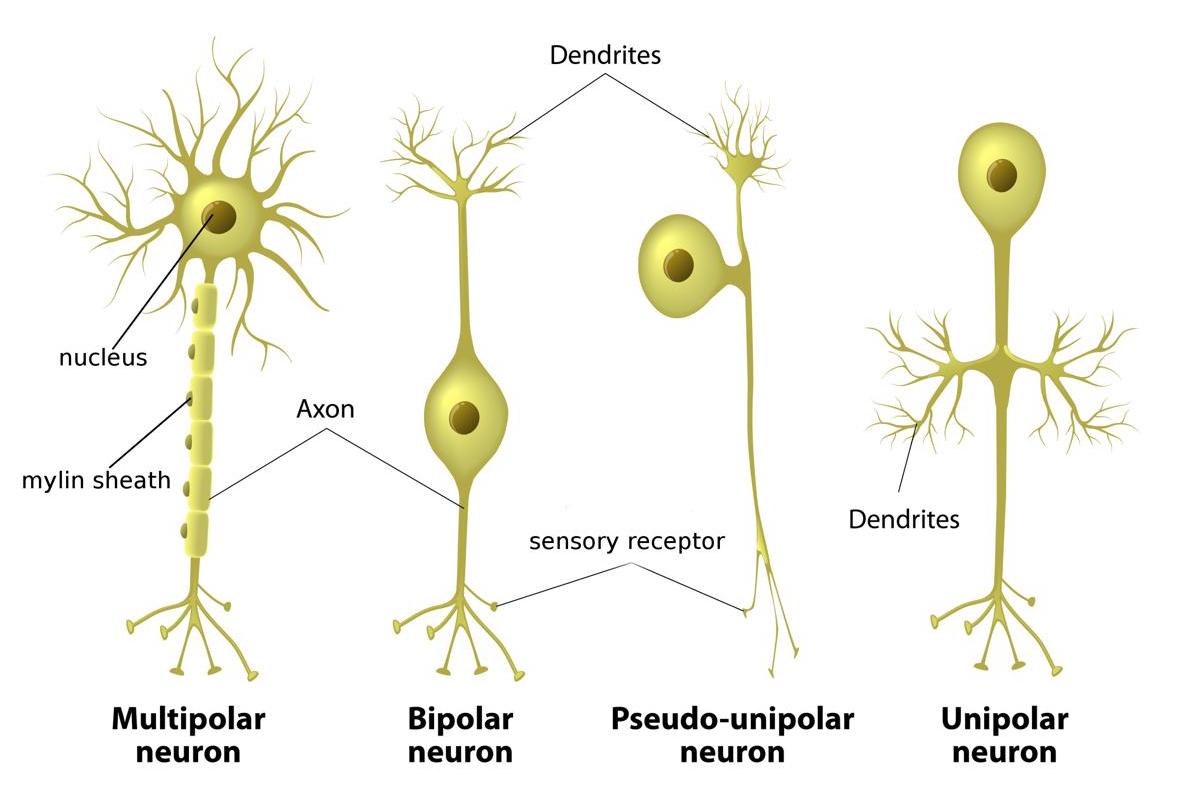}
    \caption{An Illustration of Different Neuron Structures \cite{unipolar_neuron}.}
    \label{fig:brain_neuron_structure}
\end{figure}

Neurons vary in shape and size and can be classified by their morphology and function. In this part, we will introduce a classification of the neurons by their structures, some of which are also illustrated in Figure~\ref{fig:brain_neuron_structure}.

\begin{itemize}

\item \textbf{Unipolar}: As shown in Figure~\ref{fig:brain_neuron_structure}, a unipolar neuron \cite{unipolar_neuron_wiki} is a type of neuron in which only one protoplasmic process (neurite) extends from the cell body. Unipolar neurons are common in insects, where the cell body is often located at the periphery of the brain and is electrically inactive. These cell bodies often send a single neurite into the brain; however, this neurite may ramify into a large number of branches making a very complex set of connections with other neurites, in regions of neuropil.

\item \textbf{Biploar}: A bipolar neuron or bipolar cell \cite{bipolar_neuron_wiki}, is a type of neuron which has two extensions (one axon and one dendrite). Bipolar cells are specialized sensory neurons for the transmission of special senses. As such, they are part of the sensory pathways for smell, sight, taste, hearing and vestibular functions. Common examples are the bipolar cell of the retina, the ganglia of the vestibulocochlear nerve, and the extensive use of bipolar cells to transmit efferent (motor) signals to control muscles.

\item \textbf{Multipolar}: A multipolar neuron (or multipolar neurone) \cite{multipolar_neuron_wiki} is a type of neuron that possesses a single axon and many dendrites (and dendritic branches), allowing for the integration of a great deal of information from other neurons. These processes are projections from the nerve cell body. Multipolar neurons constitute the majority of neurons in the central nervous system. They include motor neurons and interneurons/relaying neurons are most commonly found in the cortex of the brain, the spinal cord, and also in the autonomic ganglia.

\item \textbf{Anaxonic}: An anaxonic neuron \cite{anaxonic_neuron_wiki} is a neuron where the axon cannot be differentiated from the dendrites. Some sources mention that such neurons have no axons and only dendrites. They are found in the brain and retina, which act as non-spiking interneurons.

\item \textbf{Pseudounipolar}: A pseudounipolar neuron \cite{pseudounipolar_neuron_wiki} is a kind of sensory neuron in the peripheral nervous system. This neuron contains an axon that has split into two branches. These neurons have sensory receptors on skin, joints, muscles, and other parts of the body. The area of the axon that is closest to the receptor is the trigger zone for the neuron. The signal is conducted through the axon to the dorsal root ganglion's cell body, then through the axon and ending at the sensory nuclei in the dorsal column-medial lemniscus pathway of the spinal cord.

\end{itemize}

\subsection{Neuron Functional Classification}

\begin{figure}[t]
    \centering
    \includegraphics[width=0.8\textwidth]{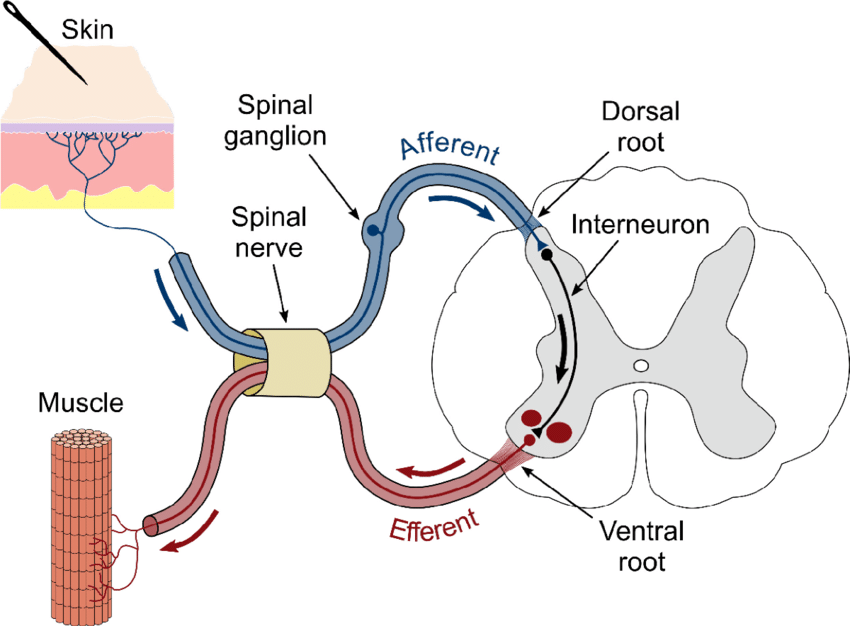}
    \caption{An Illustration of Neuron Functional Classification \cite{neuron_function_plot}.}
    \label{fig:neuron_function_plot}
\end{figure}

Meanwhile, according to the functions, neurons can be classified into three categories in another way, which are also illustrated in Figure~\ref{fig:neuron_function_plot}.
\begin{itemize}
\item \textbf{Afferent Neurons}: Afferent neurons \cite{afferent_wiki} refer to axonal projections that arrive at a particular brain region. The afferent term has a slightly different meaning in the context of the peripheral nervous system and central nervous system. In the peripheral nervous system, afferents are always from the perspective of the spinal cord (see figures). In other words, peripheral nervous system afferents are the axons of sensory neurons carrying sensory information from all over the body, into the spine. On the other hand, in the central nervous system, afferent projections can be from the perspective of any given brain region. That is, each brain region has its own unique set of afferent and efferent projections. In the context of a given brain region, afferents are arriving fibers. For instance, as shown in Figure~\ref{fig:neuron_function_plot}, the afferent neurons will pass the external stimulation received at the skin to the brain.

\item \textbf{Efferent Neurons}: Efferent neurons \cite{efferent_wiki} refer to axonal projections that exit a particular region; as opposed to afferent projections that arrive at the region. The efferent fiber is a long process projecting far from the neuron's body that carries nerve impulses away from the central nervous system toward the peripheral effector organs. A bundle of these fibers is called a motor nerve or an efferent nerve. The efferents can also be slightly different in the peripheral nervous system and central nervous system. Efferents of the peripheral nervous system are the axons of spinal cord motor neurons that carry motor-movement signals out of the spine to the muscles. Meanwhile, for the central nervous system, in the context of a given brain region, efferents are exiting fibers. For instance, in Figure~\ref{fig:neuron_function_plot}, the efferent neurons pass the brain internal nerve impulses to control the reaction of the muscle.

\item \textbf{Interneurons}: An interneuron \cite{interneuron_wiki} is a broad class of neurons found in the human body. Interneurons create neural circuits, enabling communication between sensory or motor neurons and the central nervous system. They have been found to function in reflexes, neuronal oscillations, and neurogenesis in the adult mammalian brain. Interneurons can be further broken down into two groups: local interneurons and relay interneurons. Local interneurons have short axons and form circuits with nearby neurons to analyze small pieces of information. Relay interneurons have long axons and connect circuits of neurons in one region of the brain with those in other regions. The interaction between interneurons allow the brain to perform complex functions such as learning, and decision-making. For instance, the interneuron in Figure~\ref{fig:neuron_function_plot} act as the bridge to receive the impulses from the afferent neurons, and pass the impulses tot he efferent neurons.

\end{itemize}

As indicated in Figure~\ref{fig:neuron_function_plot}, neurons in both the peripheral nervous system and the central nervous system are extensively connected to each other, where the impulses may pass from one neuron to the other neurons as the electrochemical pulse. Formally, the neuron connection points are called the synapses, and the electrochemical pulses in neuron communication is also called the action potential. In the following two sections, we will talk about these two concepts in great detail.


\section{Synapse}

\begin{figure}[t]
    \centering
    \includegraphics[width=0.8\textwidth]{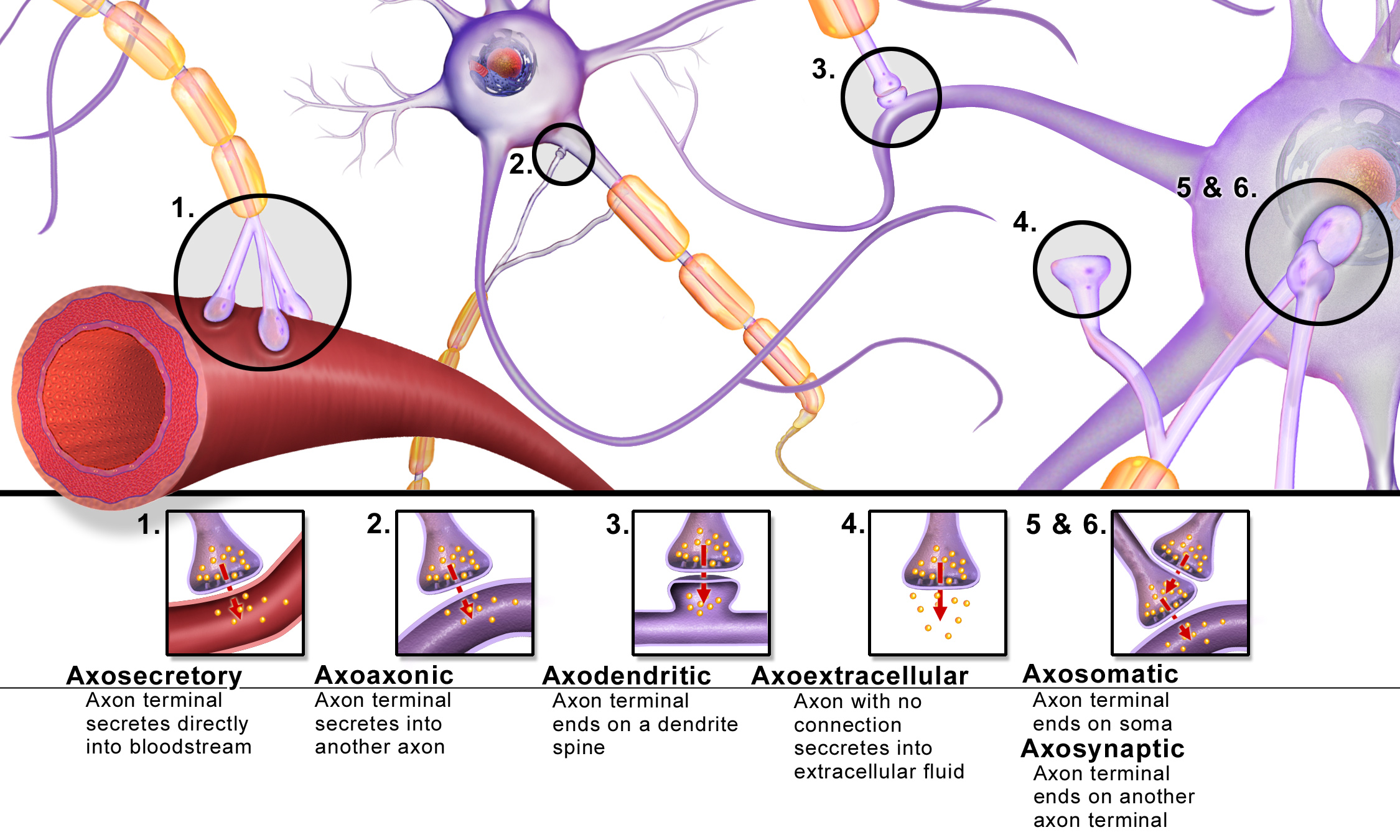}
    \caption{An Illustration of Different Synapse Interface \cite{synapse}.}
    \label{fig:synapse_interface_types}
\end{figure}

As introduced in \cite{synapse}, in the nervous system, a synapse is a structure that permits a neuron (or nerve cell) to pass an electrical or chemical signal to another neuron or to the target effector cell. Synapses are essential to neuronal function: neurons are cells that are specialized to pass signals to individual target cells, and synapses are the means by which they do so. At a synapse, the plasma membrane of the signal-passing neuron (the presynaptic neuron) comes into close apposition with the membrane of the target (postsynaptic) cell. Both the presynaptic and postsynaptic sites contain extensive arrays of a molecular machinery that link the two membranes together and carry out the signaling process. In many synapses, the presynaptic part is located on an axon and the postsynaptic part is located on a dendrite or soma. Astrocytes also exchange information with the synaptic neurons, responding to synaptic activity and, in turn, regulating neurotransmission. Synapses (at least chemical synapses) are stabilized in position by synaptic adhesion molecules (SAMs) projecting from both the pre- and post-synaptic neuron and sticking together where they overlap; SAMs may also assist in the generation and functioning of synapses.

Synapses can be classified by the type of cellular structures serving as the pre- and post-synaptic components. The vast majority of synapses in the mammalian nervous system are classical axo-dendritic synapses (axon synapsing upon a dendrite), however, as illustrated in Figure~\ref{fig:synapse_interface_types}, a variety of other arrangements exist. These include but are not limited to axo-axonic, dendro-dendritic, axo-secretory, somato-dendritic, dendro-somatic, and somato-somatic synapses. The axon can synapse onto a dendrite, onto a cell body, or onto another axon or axon terminal, as well as into the bloodstream or diffusely into the adjacent nervous tissue.

\begin{figure}[t]
    \centering
    \includegraphics[width=0.6\textwidth]{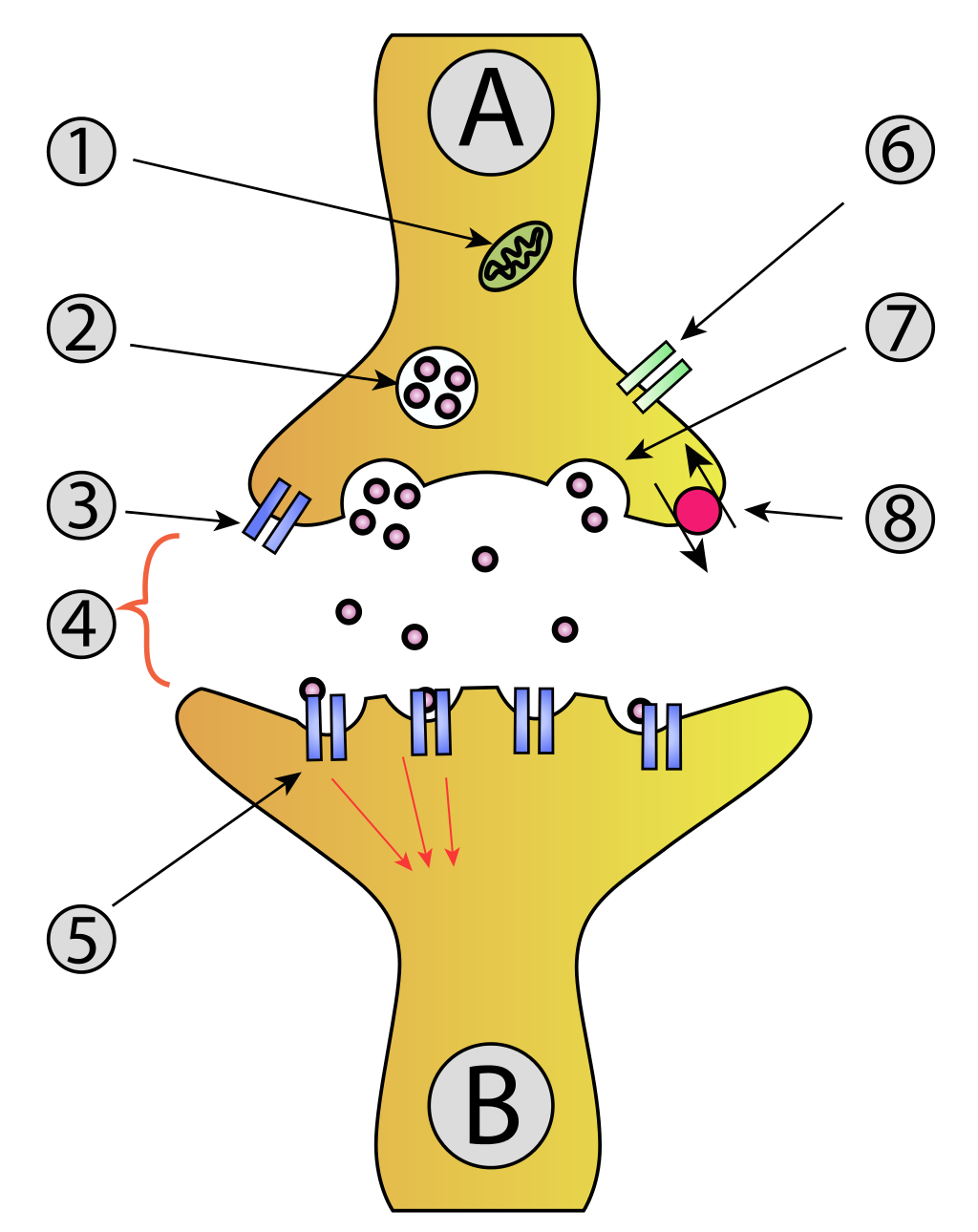}
    \caption{An Illustration of Neurotransmitters Transmission via Synapses (1: Mitochondria; 2: Synaptic vesicle with neurotransmitters; 3: Autoreceptor; 4: Synapse with neurotransmitter released (serotonin); 5: Postsynaptic receptors activated by neurotransmitter (induction of a postsynaptic potential); 6: Calcium channel; 7: Exocytosis of a vesicle; 8: Recaptured neurotransmitter) \cite{synapse}.}
    \label{fig:synapse_transmitting_process}
\end{figure}

There are two fundamentally different types of synapses:
\begin{itemize}
\item \textbf{Chemical Synapse}: In a chemical synapse, electrical activity in the presynaptic neuron is converted (via the activation of voltage-gated calcium channels) into the release of a chemical called a neurotransmitter that binds to receptors located in the plasma membrane of the postsynaptic cell. The neurotransmitter may initiate an electrical response or a secondary messenger pathway that may either excite or inhibit the postsynaptic neuron. Chemical synapses can be classified according to the neurotransmitter released: glutamatergic (often excitatory), GABAergic (often inhibitory), cholinergic (e.g. vertebrate neuromuscular junction), and adrenergic (releasing norepinephrine). Because of the complexity of receptor signal transduction, chemical synapses can have complex effects on the postsynaptic cell.

\item \textbf{Electrical Synapse}: In an electrical synapse, the presynaptic and postsynaptic cell membranes are connected by special channels called gap junctions or synaptic cleft that are capable of passing an electric current, causing voltage changes in the presynaptic cell to induce voltage changes in the postsynaptic cell. The main advantage of an electrical synapse is the rapid transfer of signals from one cell to the next.
\end{itemize}

For the chemical synapses, the process of transmitting neurotransmitters from one neuron to another neuron can be illustrated with Figure~\ref{fig:synapse_transmitting_process}. Several key structures and steps are involved, including \textit{synaptic vesicle with neurotransmitters}, \textit{autoreceptor}, \textit{synapse with neurotransmitter released (serotonin)}, \textit{postsynaptic receptors activated by neurotransmitter}, \textit{calcium channel}, \textit{exocytosis of a vesicle} and \textit{recaptured neurotransmitter}. In the following part, we will briefly introduce the \textit{synaptic vesicle} and \textit{neurotransmitter}, release and transmission of which lead to the propagation of impulses among the neurons. For the information of the other structures, readers may also refer to their wikipedia pages, which provides a detailed introduction as well.

\subsection{Synaptic Vesicle}

According to \cite{synaptic_vesicle_wiki}, in a neuron, synaptic vesicles (or neurotransmitter vesicles) store various neurotransmitters that are released at the synapse. The release is regulated by a voltage-dependent calcium channel. Vesicles are essential for propagating nerve impulses between neurons and are constantly recreated by the cell. The area in the axon that holds groups of vesicles is an axon terminal or ``terminal bouton''. Up to 130 vesicles can be released per bouton over a ten-minute period of stimulation at 0.2 Hz. In the visual cortex of the human brain, synaptic vesicles have an average diameter of 39.5 nanometers (nm) with a standard deviation of 5.1 nm.

\subsubsection{Synaptic Vesicle Composition and Proteins}

Synaptic vesicles are relatively simple because only a limited number of proteins fit into a sphere of 40 nm diameter. Purified vesicles have a protein:phospholipid ratio of 1:3 with a lipid composition of 40\% phosphatidylcholine, 32\% phosphatidylethanolamine, 12\% phosphatidylserine, 5\% phosphatidylinositol, and 10\% cholesterol.

Synaptic vesicles contain two classes of obligatory components: transport proteins involved in neurotransmitter uptake, and trafficking proteins that participate in synaptic vesicle exocytosis, endocytosis, and recycling.

\begin{itemize}
\item Transport proteins are composed of proton pumps that generate electrochemical gradients, which allow for neurotransmitter uptake, and neurotransmitter transporters that regulate the actual uptake of neurotransmitters. The necessary proton gradient is created by V-ATPase, which breaks down ATP for energy. Vesicular transporters move neurotransmitters from the cells' cytoplasm into the synaptic vesicles. Vesicular glutamate transporters, for example, sequester glutamate into vesicles by this process.
\item Trafficking proteins are more complex. They include intrinsic membrane proteins, peripherally bound proteins, and proteins such as SNAREs. These proteins do not share a characteristic that would make them identifiable as synaptic vesicle proteins, and little is known about how these proteins are specifically deposited into synaptic vesicles. Many but not all of the known synaptic vesicle proteins interact with non-vesicular proteins and are linked to specific functions.
\end{itemize}

The stoichiometry for the movement of different neurotransmitters into a vesicle is given in table~\ref{table:neurotransmitters}.

\begin{table}[t]
\caption{The Stoichiometry for the movement of Different Neurotransmitters.}
\centering
\begin{tabular}{| p{6.5cm} | c | c |}
\hline
\textbf{Neurotransmitter type(s)} & \textbf{Inward movement} & \textbf{Outward movement} \\
\hline\hline
norepinephrine, dopamine, histamine, serotonin and acetylcholine & neurotransmitter$^+$  & 2 H$^+$ \\
\hline
GABA and glycine & neurotransmitter & 1 H$^+$ \\
\hline
glutamate &neurotransmitter$^-$ + cl$^-$  & 1 H$^+$\\
\hline
\end{tabular}
\label{table:neurotransmitters}
\end{table}

\subsubsection{Synaptic Vesicle Pools}

\begin{figure}[t]
    \centering
    \includegraphics[width=0.6\textwidth]{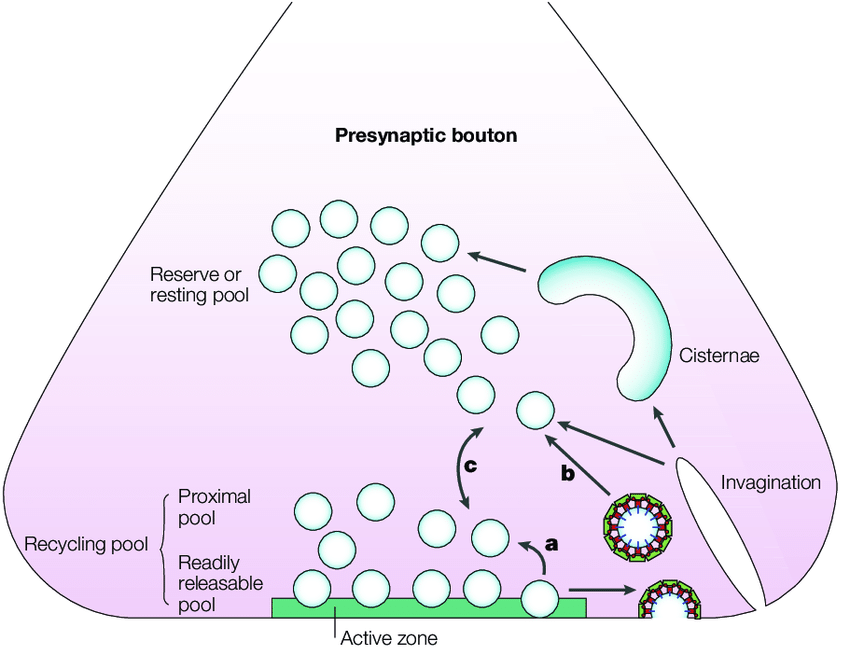}
    \caption{Synaptic-vesicle pools in presynaptic boutons. \cite{vesicle_pools}.}
    \label{fig:vesicle_pools}
\end{figure}

As introduced in \cite{vesicle_pools}, according to their availability for neurotransmitter release and/or according to their localization in the nerve terminal, various pools of synaptic vesicles can be distinguished, which are also illustrated in Figure~\ref{fig:vesicle_pools}. Vesicles of the rapidly recycling pool participate actively in exocytosis under conditions of physiological stimulation, whereas the reserve or resting pool includes synaptic vesicles that are activated only in response to strong synaptic stimulation. Vesicles of the recycling pool can be further divided into the readily releasable pool of synaptic vesicles, which are docked and primed and can exocytose their content within milliseconds after stimulation, and the proximal pool, which can rapidly replace vesicles that have released their content. At a typical brain synapse, the rapidly recycling pool might consist of about 25 vesicles, of which 5-8 are readily releasable. Different modes of coupling of exocytosis and endocytosis can feed into different synaptic vesicle pools. The ``kiss-and-run'' mode is thought to feed into the recycling pool Figure~\ref{fig:vesicle_pools}-(a), whereas clathrin-mediated endocytosis feeds into the reserve pool Figure~\ref{fig:vesicle_pools}-(b). For example, at the neuromuscular junction of Drosophila melanogaster null-mutants for the endophilin A gene, in which pathway b is essentially blocked, about 15-20\% of synaptic vesicles are still cycling. These authors propose that these vesicles reflect the recycling synaptic vesicle pool, which is maintained by an endophilin-independent kiss-and-run mode. Vesicles of the reserve or resting pool might replace aged cycling synaptic vesicles or refill the recycling pool during extensive synaptic stimulation Figure~\ref{fig:vesicle_pools}-(c). A similar functional organization of excitatory vesicles has been described for adrenal chromaffin cells.

\subsubsection{The Synaptic Vesicle Cycle}

\begin{figure}[t]
    \centering
    \includegraphics[width=0.6\textwidth]{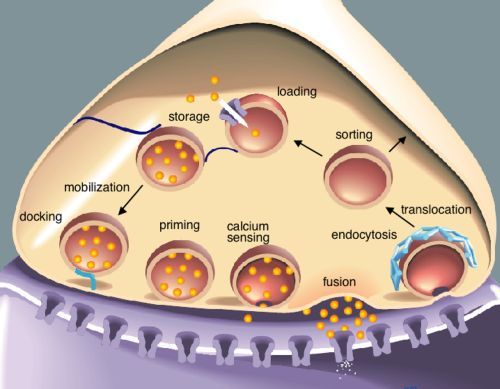}
    \caption{The synaptic vesicle cycle. \cite{synaptic_vesicle_cycle}.}
    \label{fig:vesicle_cycle}
\end{figure}

According to \cite{synapse}, as illustrated in Figure~\ref{fig:vesicle_cycle}, the events of the synaptic vesicle cycle can be divided into a few key steps:

\begin{enumerate}
\item \textbf{Trafficking to the synapse}: Synaptic vesicle components are initially trafficked to the synapse using members of the kinesin motor family. In Caenorhabditis elegans the major motor for synaptic vesicles is UNC-104. There is also evidence that other proteins such as UNC-16/Sunday Driver regulate the use of motors for transport of synaptic vesicles.

\item \textbf{Transmitter loading}: Once at the synapse, synaptic vesicles are loaded with a neurotransmitter. Loading of transmitter is an active process requiring a neurotransmitter transporter and a proton pump ATPase that provides an electrochemical gradient. These transporters are selective for different classes of transmitters. Characterization of unc-17 and unc-47, which encode the vesicular acetylcholine transporter and vesicular GABA transporter have been described to date.

\item \textbf{Docking}: The loaded synaptic vesicles must dock near release sites, however docking is a step of the cycle that we know little about. Many proteins on synaptic vesicles and at release sites have been identified, however none of the identified protein interactions between the vesicle proteins and release site proteins can account for the docking phase of the cycle. Mutants in rab-3 and unc-18 alter vesicle docking or vesicle organization at release sites, but they do not completely disrupt docking. SNARE proteins, do not appear to be involved in the docking step of the cycle.

\item \textbf{Priming}: After the synaptic vesicles initially dock, they must be primed before they can begin fusion. Priming prepares the synaptic vesicle so that they are able to fuse rapidly in response to a calcium influx. This priming step is thought to involve the formation of partially assembled SNARE complexes. The proteins Munc13, RIM, and RIM-BP participate in this event. Munc13 is thought to stimulate the change of the t-SNARE syntaxin from a closed conformation to an open conformation, which stimulates the assembly of v-SNARE /t-SNARE complexes. RIM also appears to regulate priming, but is not essential for the step.

\item \textbf{Fusion}: Primed vesicles fuse very quickly in response to calcium elevations in the cytoplasm. This fusion event is thought to be mediated directly by the SNAREs and driven by the energy provided from SNARE assembly. The calcium-sensing trigger for this event is the calcium-binding synaptic vesicle protein synaptotagmin. The ability of SNAREs to mediate fusion in a calcium-dependent manner recently has been reconstituted in vitro. Consistent with SNAREs being essential for the fusion process, v-SNARE and t-SNARE mutants of C. elegans are lethal. Similarly, mutants in Drosophila and knockouts in mice indicate that these SNARES play a critical role in synaptic exocytosis.

\item \textbf{Endocytosis}: This accounts for the re-uptake of synaptic vesicles in the full contact fusion model. However, other studies have been compiling evidence suggesting that this type of fusion and endocytosis is not always the case.
\end{enumerate}

\subsubsection{Vesicle Recycling}

\begin{figure}[t]
    \centering
    \includegraphics[width=0.6\textwidth]{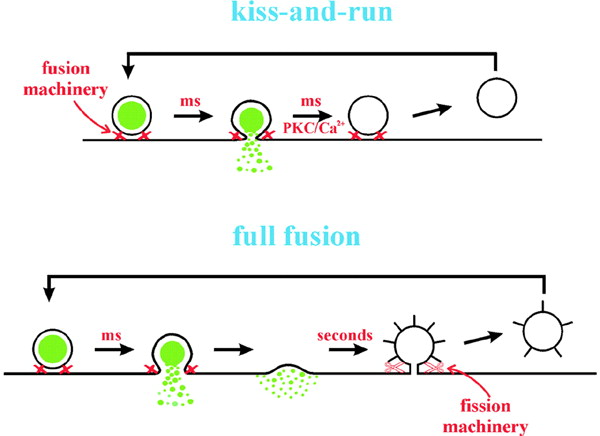}
    \caption{The Synaptic Vesicle Recycle. \cite{synaptic_vesicle_recycle}.}
    \label{fig:vesicle_recycle}
\end{figure}

The synaptic vesicle will be recycled after releasing the neurotransmitters. Two leading mechanisms of action are thought to be responsible for synaptic vesicle recycling: full collapse fusion and the ``kiss-and-run'' method. Both mechanisms begin with the formation of the synaptic pore that releases transmitter to the extracellular space. After release of the neurotransmitter, the pore can either dilate fully so that the vesicle collapses completely into the synaptic membrane, or it can close rapidly and pinch off the membrane to generate kiss-and-run fusion.

\begin{itemize}
\item \textbf{Full collapse fusion}: It has been shown that periods of intense stimulation at neural synapses deplete vesicle count as well as increase cellular capacitance and surface area. This indicates that after synaptic vesicles release their neurotransmitter payload, they merge with and become part of, the cellular membrane. After tagging synaptic vesicles with HRP (horseradish peroxidase), Heuser and Reese found that portions of the cellular membrane at the frog neuromuscular junction were taken up by the cell and converted back into synaptic vesicles. Studies suggest that the entire cycle of exocytosis, retrieval, and reformation of the synaptic vesicles requires less than 1 minute.

In full collapse fusion, the synaptic vesicle merges and becomes incorporated into the cell membrane. The formation of the new membrane is a protein mediated process and can only occur under certain conditions. After an action potential, Ca$^{2+}$ floods to the presynaptic membrane. Ca$^{2+}$ binds to specific proteins in the cytoplasm, one of which is synaptotagmin, which in turn trigger the complete fusion of the synaptic vesicle with the cellular membrane. This complete fusion of the pore is assisted by SNARE proteins. This large family of proteins mediate docking of synaptic vesicles in an ATP-dependent manner. With the help of synaptobrevin on the synaptic vesicle, the t-SNARE complex on the membrane, made up of syntaxin and SNAP-25, can dock, prime, and fuse the synaptic vesicle into the membrane.

The mechanism behind full collapse fusion has been shown to be the target of the botulinum and tetanus toxins. The botulinum toxin has protease activity which degrades the SNAP-25 protein. The SNAP-25 protein is required for vesicle fusion that releases neurotransmitters, in particular acetylcholine. Botulinum toxin essentially cleaves these SNARE proteins, and in doing so, prevents synaptic vesicles from fusing with the cellular synaptic membrane and releasing their neurotransmitters. Tetanus toxin follows a similar pathway, but instead attacks the protein synaptobrevin on the synaptic vesicle. In turn, these neurotoxins prevent synaptic vesicles from completing full collapse fusion. Without this mechanism in effect, muscle spasms, paralysis, and death can occur.

\item \textbf{``Kiss-and-run''}: The second mechanism by which synaptic vesicles are recycled is known as kiss-and-run fusion. In this case, the synaptic vesicle ``kisses'' the cellular membrane, opening a small pore for its neurotransmitter payload to be released through, then closes the pore and is recycled back into the cell. The kiss-and-run mechanism has been a hotly debated topic. Its effects have been observed and recorded; however the reason behind its use as opposed to full collapse fusion is still being explored. It has been speculated that kiss-and-run is often employed to conserve scarce vesicular resources as well as being utilized to respond to high-frequency inputs. Experiments have shown that kiss-and-run events do occur. First observed by Katz and del Castillo, it was later observed that the kiss-and-run mechanism was different from full collapse fusion in that cellular capacitance did not increase in kiss-and-run events. This reinforces the idea of a kiss-and-run fashion, the synaptic vesicle releases its payload and then separates from the membrane.
\end{itemize}

Cells thus appear to have at least two mechanisms to follow for membrane recycling. Under certain conditions, cells can switch from one mechanism to the other. Slow, conventional, full collapse fusion predominates the synaptic membrane when Ca$^{2+}$ levels are low, and the fast kiss-and-run mechanism is followed when Ca$^{2+}$ levels are high.

\subsection{Neurotransmitter}

As introduced in \cite{neurotransmitter_wiki}, neurotransmitters are endogenous chemicals that enable neurotransmission. It is a type of chemical messenger which transmits signals across a chemical synapse, such as a neuromuscular junction, from one neuron (nerve cell) to another ``target'' neuron, muscle cell, or gland cell. Neurotransmitters are released from synaptic vesicles in synapses into the synaptic cleft, where they are received by neurotransmitter receptors on the target cells. Many neurotransmitters are synthesized from simple and plentiful precursors such as amino acids, which are readily available from the diet and only require a small number of biosynthetic steps for conversion. Neurotransmitters play a major role in shaping everyday life and functions. Their exact numbers are unknown, but more than 200 chemical messengers have been uniquely identified.

Neurotransmitters are stored in synaptic vesicles, clustered close to the cell membrane at the axon terminal of the presynaptic neuron. Neurotransmitters are released into and diffuse across the synaptic cleft, where they bind to specific receptors on the membrane of the postsynaptic neuron. Most neurotransmitters are about the size of a single amino acid; however, some neurotransmitters may be the size of larger proteins or peptides. A released neurotransmitter is typically available in the synaptic cleft for a short time before it is metabolized by enzymes, pulled back into the presynaptic neuron through reuptake, or bound to a postsynaptic receptor. Nevertheless, short-term exposure of the receptor to a neurotransmitter is typically sufficient for causing a postsynaptic response by way of synaptic transmission.

In response to a threshold action potential or graded electrical potential, a neurotransmitter is released at the presynaptic terminal. Low level ``baseline'' release also occurs without electrical stimulation. The released neurotransmitter may then move across the synapse to be detected by and bind with receptors in the postsynaptic neuron. Binding of neurotransmitters may influence the postsynaptic neuron in either an inhibitory or excitatory way. This neuron may be connected to many more neurons, and if the total of excitatory influences are greater than those of inhibitory influences, the neuron will also ``fire''. Ultimately it will create a new action potential at its axon hillock to release neurotransmitters and pass on the information to yet another neighboring neuron.

\subsubsection{Neurotransmitter Identification}

There are four main criteria for identifying neurotransmitters:
\begin{itemize}
\item The chemical must be synthesized in the neuron or otherwise be present in it.
\item When the neuron is active, the chemical must be released and produce a response in some target.
\item The same response must be obtained when the chemical is experimentally placed on the target.
\item A mechanism must exist for removing the chemical from its site of activation after its work is done.
\end{itemize}

However, given advances in pharmacology, genetics, and chemical neuroanatomy, the term ``neurotransmitter'' can be applied to chemicals that:

\begin{itemize}
\item Carry messages between neurons via influence on the postsynaptic membrane.
\item Have little or no effect on membrane voltage, but have a common carrying function such as changing the structure of the synapse.
\item Communicate by sending reverse-direction messages that affect the release or reuptake of transmitters.
\end{itemize}

The anatomical localization of neurotransmitters is typically determined using immunocytochemical techniques, which identify the location of either the transmitter substances themselves, or of the enzymes that are involved in their synthesis. Immunocytochemical techniques have also revealed that many transmitters, particularly the neuropeptides, are co-localized, that is, one neuron may release more than one transmitter from its synaptic terminal. Various techniques and experiments such as staining, stimulating, and collecting can be used to identify neurotransmitters throughout the central nervous system.

\subsubsection{Neurotransmitter Types}

\begin{figure}[t]
    \centering
    \includegraphics[width=1.0\textwidth]{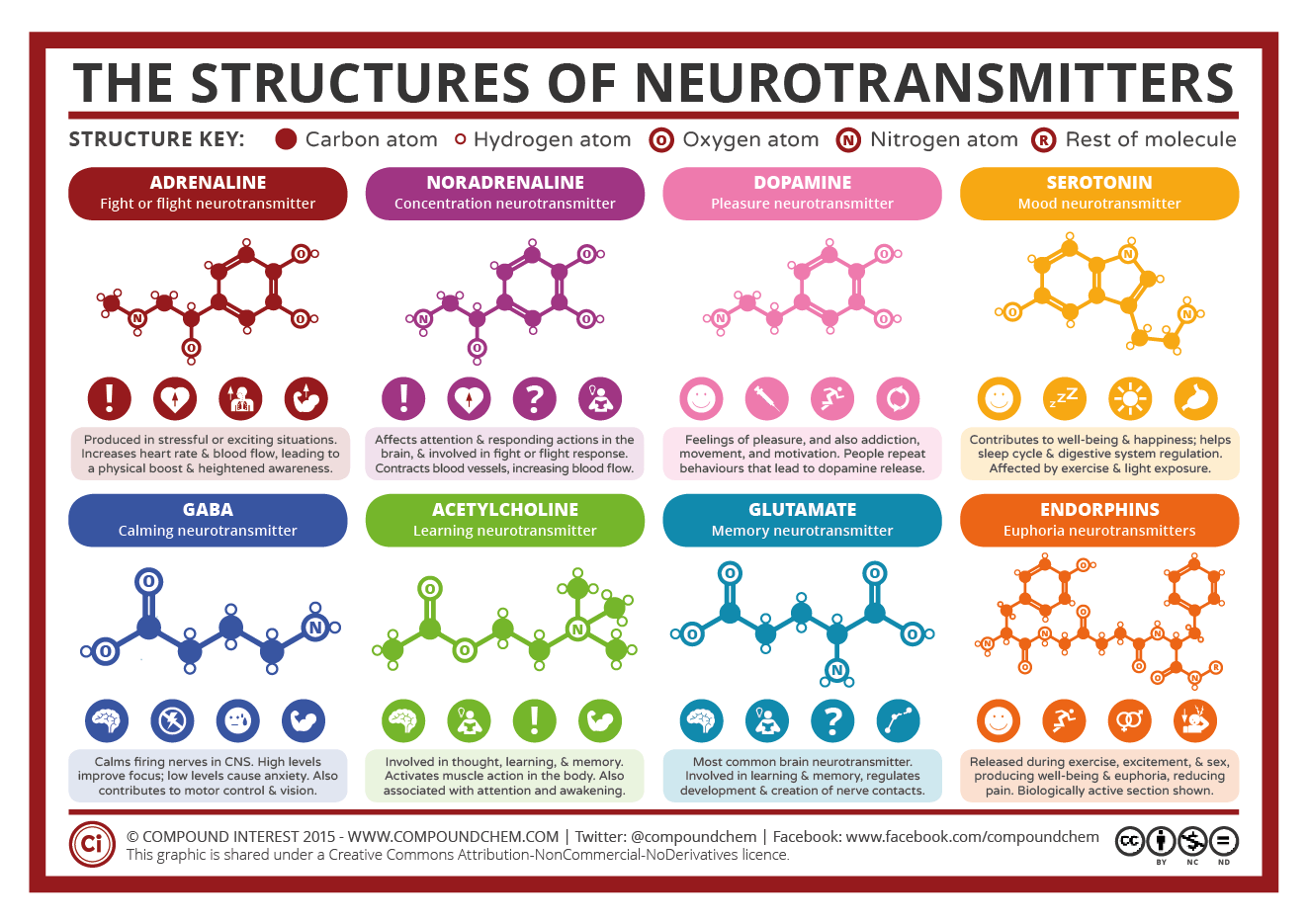}
    \caption{An Illustration of Several Major Neurotransmitters \cite{neurotransmitter_examples}.}
    \label{fig:neurotransmitter}
\end{figure}

There are many different ways to classify neurotransmitters. Dividing them into amino acids, peptides, and monoamines is sufficient for some classification purposes.

Major neurotransmitters:
\begin{itemize}
\item \textbf{Amino acids}: glutamate, aspartate, D-serine, $\gamma$-aminobutyric acid (GABA), glycine
\item \textbf{Gasotransmitters}: nitric oxide (NO), carbon monoxide (CO), hydrogen sulfide (H2S)
\item \textbf{Monoamines}: dopamine (DA), norepinephrine (noradrenaline; NE, NA), epinephrine (adrenaline), histamine, serotonin (SER, 5-HT)
\item \textbf{Trace amines}: phenethylamine, N-methylphenethylamine, tyramine, 3-iodothyronamine, octopamine, tryptamine, etc.
\item \textbf{Peptides}: oxytocin, somatostatin, substance P, cocaine and amphetamine regulated transcript, opioid peptides
\item \textbf{Purines}: adenosine triphosphate (ATP), adenosine
\item \textbf{Catecholamines}: dopamine, norepinephrine (noradrenaline), epinephrine (adrenaline)
\item \textbf{Others}: acetylcholine (ACh), anandamide, etc.
\end{itemize}

The molecular structure, formula and function of several neurotransmitters are also shown in Figure~\ref{fig:neurotransmitter}. In addition, over 50 neuroactive peptides have been found, and new ones are discovered regularly. Many of these are ``co-released'' along with a small-molecule transmitter. Nevertheless, in some cases a peptide is the primary transmitter at a synapse. $\beta$-endorphin is a relatively well-known example of a peptide neurotransmitter because it engages in highly specific interactions with opioid receptors in the central nervous system.

Single ions (such as synaptically released zinc) are also considered neurotransmitters by some, as well as some gaseous molecules such as nitric oxide (NO), carbon monoxide (CO), and hydrogen sulfide (H2S). The gases are produced in the neural cytoplasm and are immediately diffused through the cell membrane into the extracellular fluid and into nearby cells to stimulate production of second messengers. Soluble gas neurotransmitters are difficult to study, as they act rapidly and are immediately broken down, existing for only a few seconds.

The most prevalent transmitter is glutamate, which is excitatory at well over 90\% of the synapses in the human brain. The next most prevalent is Gamma-Aminobutyric Acid, or GABA, which is inhibitory at more than 90\% of the synapses that do not use glutamate. Although other transmitters are used in fewer synapses, they may be very important functionally: the great majority of psychoactive drugs exert their effects by altering the actions of some neurotransmitter systems, often acting through transmitters other than glutamate or GABA. Addictive drugs such as cocaine and amphetamines exert their effects primarily on the dopamine system. The addictive opiate drugs exert their effects primarily as functional analogs of opioid peptides, which, in turn, regulate dopamine levels. An incomplete list of neurotransmitters, peptides, and gaseous signaling molecules is also provided at the wiki-page \cite{neurotransmitter_wiki}.


\section{Neuron Membrane Potential and Action Potential}\label{sec:action_potential}

In this section, we will talk about the neuron membrane potential, which is key to understand how the neurons are activated to transmit impulses among the other neurons in the nervous system as mentioned before in both this and the previous tutorial article \cite{zhang2019secrets}. Formally, membrane potential (also transmembrane potential or membrane voltage) is the difference in electric potential between the interior and the exterior of a biological cell. According to \cite{membrane_potential}, with respect to the exterior of the cell, typical values of membrane potential, normally given in millivolts, range from -40 mV to -80 mV.

All animal cells are surrounded by a membrane composed of a lipid bilayer with proteins embedded in it. The membrane serves as both an insulator and a diffusion barrier to the movement of ions. As illustrated in Figure~\ref{fig:brain_cell_membrane} in Section~\ref{subsec:neuron_membrane}, transmembrane proteins, also known as ion transporter or ion pump proteins, actively push ions across the membrane and establish concentration gradients across the membrane, and ion channels allow ions to move across the membrane down those concentration gradients. Ion pumps and ion channels are electrically equivalent to a set of batteries and resistors inserted in the membrane, and therefore create a voltage between the two sides of the membrane.

Almost all plasma membranes have an electrical potential across them, with the inside usually negative with respect to the outside. The membrane potential has two basic functions. First, it allows a cell to function as a battery, providing power to operate a variety of ``molecular devices'' embedded in the membrane. Second, in electrically excitable cells such as neurons and muscle cells, it is used for transmitting signals between different parts of a cell. Signals are generated by opening or closing of ion channels at one point in the membrane, producing a local change in the membrane potential. This change in the electric field can be quickly affected by either adjacent or more distant ion channels in the membrane. Those ion channels can then open or close as a result of the potential change, reproducing the signal.

In non-excitable cells, and in excitable cells in their baseline states, the membrane potential is held at a relatively stable value, called the resting potential. For neurons, typical values of the resting potential range from -70 to -80 millivolts; that is, the interior of a cell has a negative baseline voltage of a bit less than one-tenth of a volt. The opening and closing of ion channels can induce a departure from the resting potential. This is called a depolarization if the interior voltage becomes less negative (say from -70 mV to -60 mV), or a hyperpolarization if the interior voltage becomes more negative (say from -70 mV to -80 mV). In excitable cells, a sufficiently large depolarization can evoke an action potential, in which the membrane potential changes rapidly and significantly for a short time (on the order of 1 to 100 milliseconds), often reversing its polarity. Action potentials are generated by the activation of certain voltage-gated ion channels.

In neurons, the factors that influence the membrane potential are diverse. They include numerous types of ion channels, some of which are chemically gated and some of which are voltage-gated. Because voltage-gated ion channels are controlled by the membrane potential, while the membrane potential itself is influenced by these same ion channels, feedback loops that allow for complex temporal dynamics arise, including oscillations and regenerative events such as action potentials. 

In this section, we will first introduce the neuron membrane structure together with the ion channels inserted in the membrane. Then we will talk about the different membrane potentials, indicating different state of the neuron. At the end, we will describe the complete working mechanism of a neuron to fire an impulse by associating the membrane potential changes with the open and close of the membrane ion channels.

\subsection{Neuron Membrane Potential}

\subsubsection{Membrane Structure}

\begin{figure}[t]
    \centering
    \includegraphics[width=1.0\textwidth]{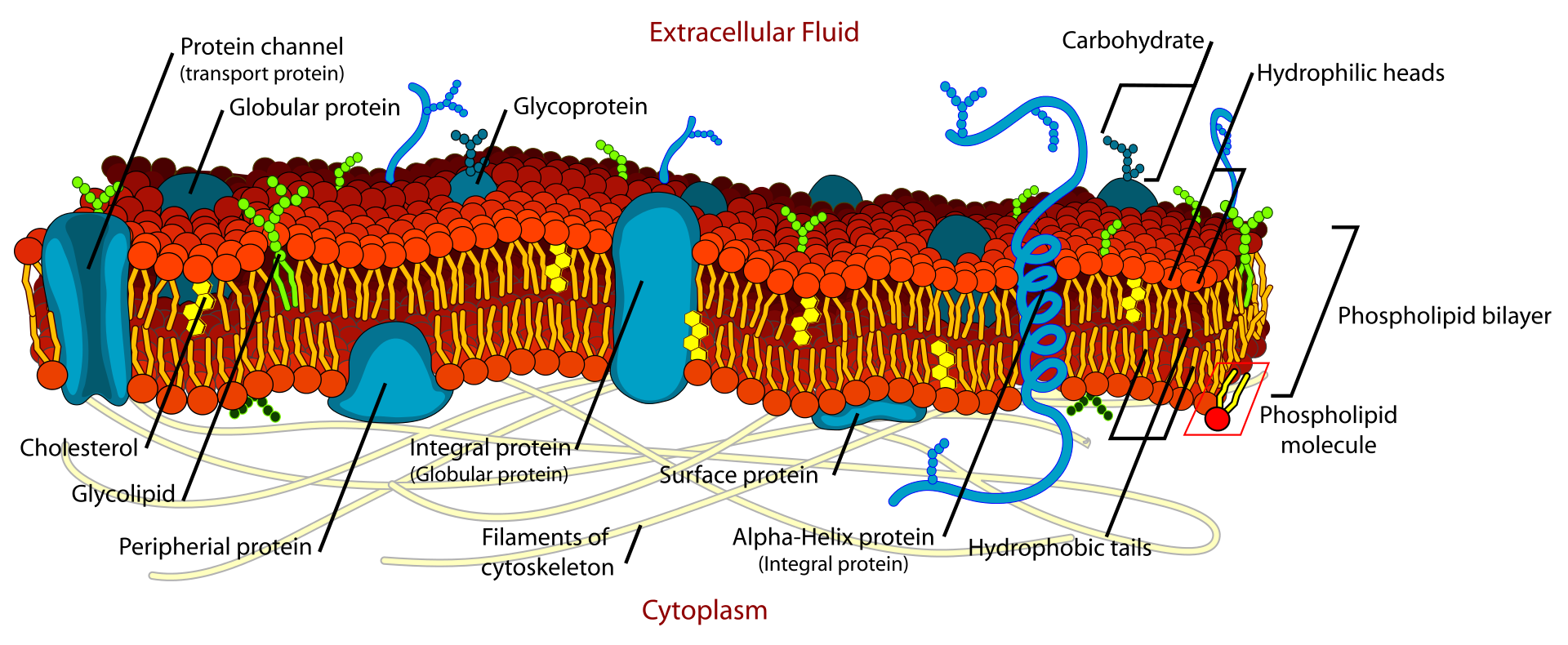}
    \caption{An Illustration of Membrane Structure \cite{membrane_potential}.}
    \label{fig:membrane_structure}
\end{figure}

According to \cite{membrane_potential}, every animal cell is enclosed in a plasma membrane, which has the structure of a lipid bilayer with many types of large molecules embedded in it, as illustrated in Figure~\ref{fig:membrane_structure}. Because it is made of lipid molecules, the plasma membrane intrinsically has a high electrical resistivity, in other words a low intrinsic permeability to ions. However, some of the molecules embedded in the membrane are capable either of actively transporting ions from one side of the membrane to the other or of providing channels through which they can move.

In electrical terminology, the plasma membrane functions as a combined resistor and capacitor. Resistance arises from the fact that the membrane impedes the movement of charges across it. Capacitance arises from the fact that the lipid bilayer is so thin that an accumulation of charged particles on one side gives rise to an electrical force that pulls oppositely charged particles toward the other side. The capacitance of the membrane is relatively unaffected by the molecules that are embedded in it, so it has a more or less invariant value estimated at about $2\mu F/{cm}^2$ (the total capacitance of a patch of membrane is proportional to its area). The conductance of a pure lipid bilayer is so low, on the other hand, that in biological situations it is always dominated by the conductance of alternative pathways provided by embedded molecules. Thus, the capacitance of the membrane is more or less fixed, but the resistance is highly variable.

The thickness of a plasma membrane is estimated to be about 7-8 nanometers. Because the membrane is so thin, it does not take a very large transmembrane voltage to create a strong electric field within it. Typical membrane potentials in animal cells are on the order of 100 millivolts (that is, one tenth of a volt), but calculations show that this generates an electric field close to the maximum that the membrane can sustain-it has been calculated that a voltage difference much larger than 200 millivolts could cause dielectric breakdown, that is, arcing across the membrane.

\subsubsection{Ion Channel}

\begin{figure}[t]
    \centering
    \includegraphics[width=0.6\textwidth]{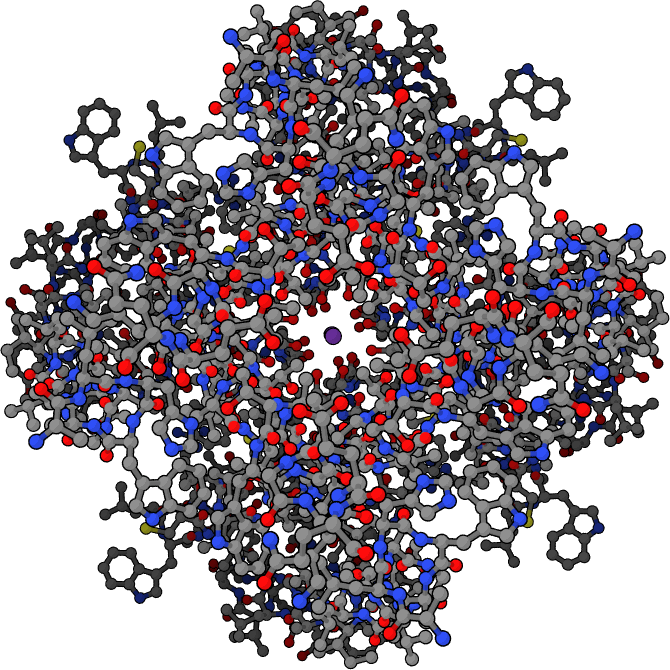}
    \caption{An Illustration of an Open Ion-Channel Protein Structure \cite{membrane_potential} (the purple ion in the center denotes a passing potassium ion).}
    \label{fig:membrane_channel_protein}
\end{figure}

Ion channels are integral membrane proteins with a pore through which ions can travel between extracellular space and cell interior. Most channels are specific (selective) for one ion; for example, most potassium channels are characterized by 1000:1 selectivity ratio for potassium over sodium, though potassium and sodium ions have the same charge and differ only slightly in their radius. The channel pore is typically so small that ions must pass through it in single-file order. Channel pores can be either open or closed for ion passage, although a number of channels demonstrate various sub-conductance levels. When a channel is open, ions permeate through the channel pore down the transmembrane concentration gradient for that particular ion. Rate of ionic flow through the channel, i.e. single-channel current amplitude, is determined by the maximum channel conductance and electrochemical driving force for that ion, which is the difference between the instantaneous value of the membrane potential and the value of the reversal potential.

A channel may have several different states (corresponding to different conformations of the protein), but each such state is either open or closed. In general, closed states correspond either to a contraction of the pore-making it impassable to the ion-or to a separate part of the protein, stoppering the pore. For example, the voltage-dependent sodium channel undergoes inactivation, in which a portion of the protein swings into the pore, sealing it. This inactivation shuts off the sodium current and plays a critical role in the action potential.

Ion channels can be classified by how they respond to their environment. For example, the ion channels involved in the action potential are voltage-sensitive channels; they open and close in response to the voltage across the membrane. Ligand-gated channels form another important class; these ion channels open and close in response to the binding of a ligand molecule, such as a neurotransmitter. Other ion channels open and close with mechanical forces. Still other ion channels-such as those of sensory neurons-open and close in response to other stimuli, such as light, temperature or pressure.

\begin{itemize}

\item \textbf{Leakage channels}: Leakage channels are the simplest type of ion channel, in that their permeability is more or less constant. The types of leakage channels that have the greatest significance in neurons are potassium and chloride channels. It should be noted that even these are not perfectly constant in their properties: First, most of them are voltage-dependent in the sense that they conduct better in one direction than the other (in other words, they are rectifiers); second, some of them are capable of being shut off by chemical ligands even though they do not require ligands in order to operate. An example of the leakage channel is illustrated in Figure~\ref{fig:leakage_channel}.
\begin{figure}[H]
    \centering
    \includegraphics[width=0.8\textwidth]{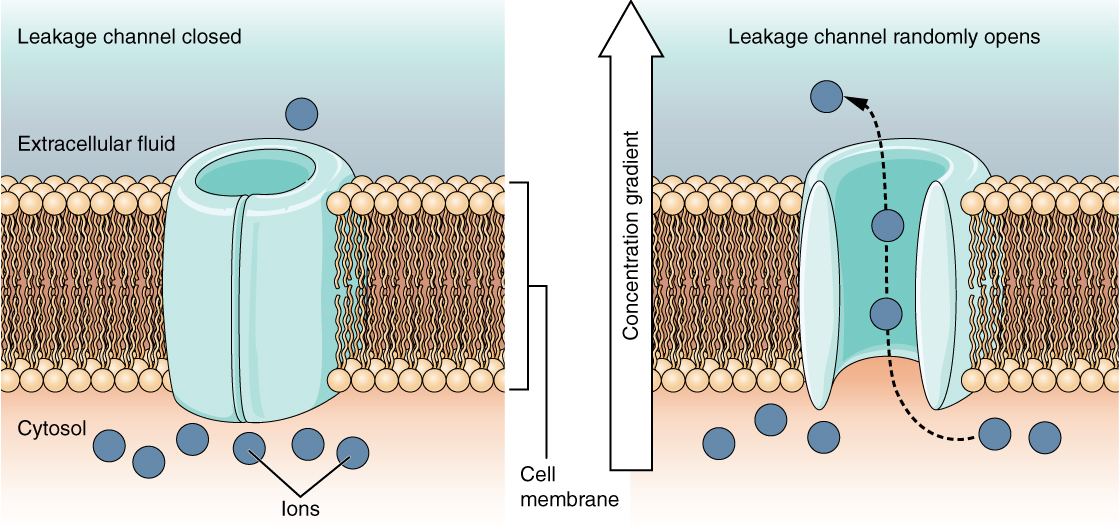}
    \caption{An Illustration of the Leakage Channel \cite{ion_gated_channel}.}
    \label{fig:leakage_channel}
\end{figure}

\item \textbf{Ligand-gated channels}: Ligand-gated ion channels are channels whose permeability is greatly increased when some type of chemical ligand binds to the protein structure. Animal cells contain hundreds, if not thousands, of types of these. A large subset function as neurotransmitter receptors-they occur at postsynaptic sites, and the chemical ligand that gates them is released by the presynaptic axon terminal. One example of this type is the AMPA receptor, a receptor for the neurotransmitter glutamate that when activated allows passage of sodium and potassium ions. Another example is the GABAA receptor, a receptor for the neurotransmitter GABA that when activated allows passage of chloride ions.

Neurotransmitter receptors are activated by ligands that appear in the extracellular area as illustrated in Figure~\ref{fig:ligand-gated_channel}, but there are other types of ligand-gated channels that are controlled by interactions on the intracellular side.

\begin{figure}[H]
    \centering
    \includegraphics[width=0.8\textwidth]{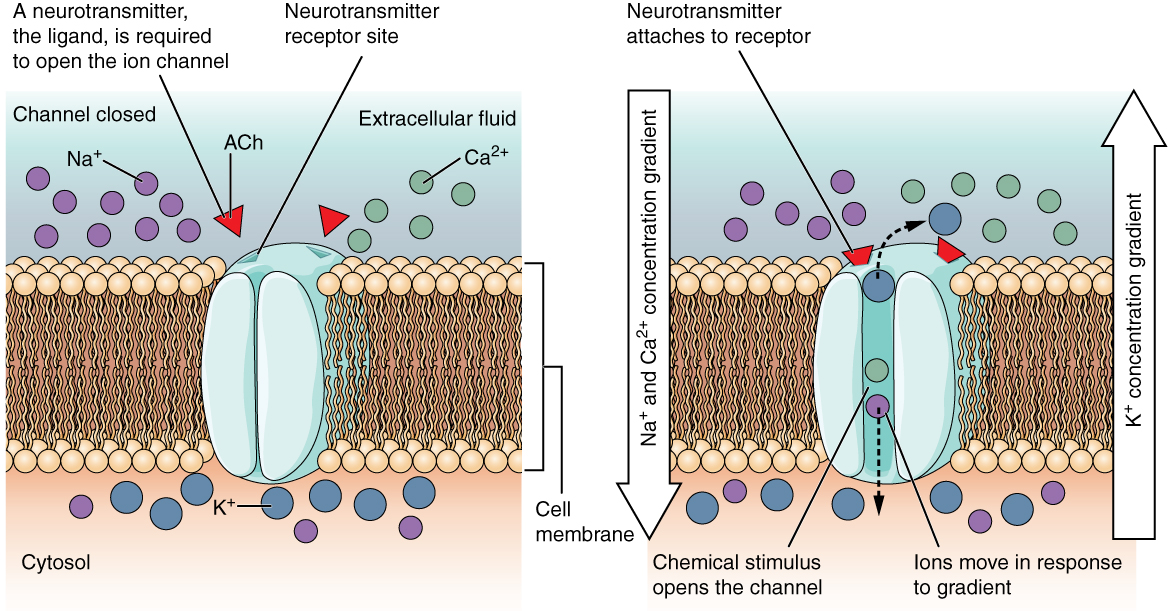}
    \caption{An Illustration of the Ligand-gated Channel \cite{ion_gated_channel}.}
    \label{fig:ligand-gated_channel}
\end{figure}

\item \textbf{Voltage-dependent channels}: Voltage-gated ion channels, also known as voltage dependent ion channels, are channels whose permeability is influenced by the membrane potential. They form another very large group, with each member having a particular ion selectivity and a particular voltage dependence. Many are also time-dependent-in other words, they do not respond immediately to a voltage change but only after a delay.

One of the most important members of this group is a type of voltage-gated sodium channel that underlies action potentials-these are sometimes called Hodgkin-Huxley sodium channels because they were initially characterized by Alan Lloyd Hodgkin and Andrew Huxley in their Nobel Prize-winning studies of the physiology of the action potential. The channel is closed at the resting voltage level, but opens abruptly when the voltage exceeds a certain threshold, allowing a large influx of sodium ions that produces a very rapid change in the membrane potential. Recovery from an action potential is partly dependent on a type of voltage-gated potassium channel that is closed at the resting voltage level but opens as a consequence of the large voltage change produced during the action potential.

As illustrated in Figure~\ref{fig:voltage_gated_channel-gated_channel}, we show a Ca$^{2+}$ channel which remains closed when the membrane voltage is -70mV and opens when the membrane voltage changes to -50mV.
\begin{figure}[H]
\vspace{-10pt}
    \centering
    \includegraphics[width=0.8\textwidth]{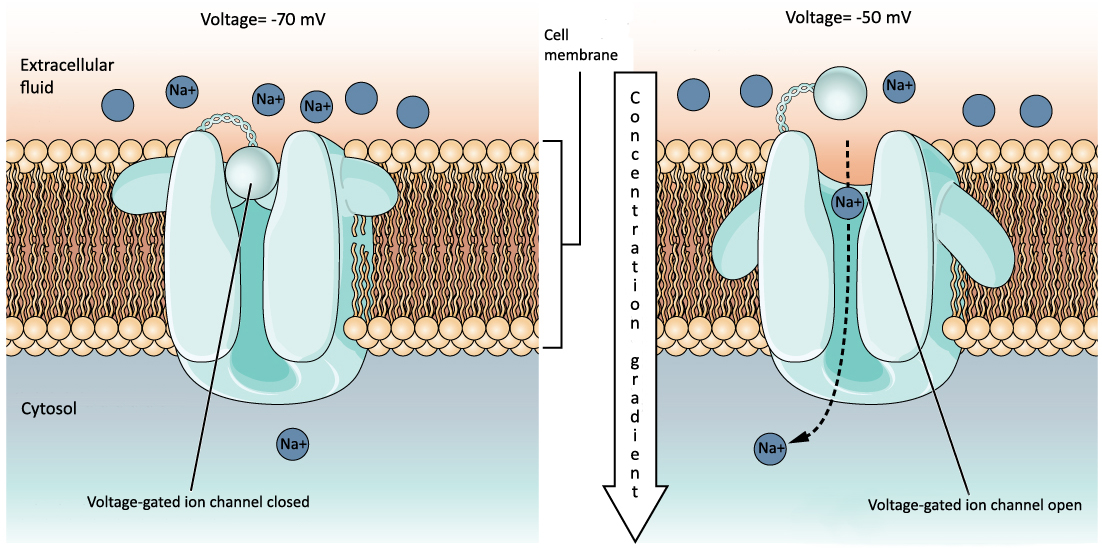}
    \caption{An Illustration of the Voltage-gated Channel \cite{ion_gated_channel}.}
    \label{fig:voltage_gated_channel-gated_channel}
\end{figure}

\item {Mechanically gated channels}: A mechanically gated channel opens because of a physical distortion of the cell membrane. Many channels associated with the sense of touch (somatosensation) are mechanically gated. For example, as pressure is applied to the skin, these channels open and allow ions to enter the cell. Similar to this type of channel would be the channel that opens on the basis of temperature changes, as in testing the water in the shower.

As illustrated in Figure~\ref{fig:mechanically_gated_channel-gated_channel}, we show an example of the mechanically gated channel. When a mechanical change occurs in the surrounding tissue, such as pressure or touch, the channel is physically opened. Thermoreceptors work on a similar principle. When the local tissue temperature changes, the protein reacts by physically opening the channel.

\begin{figure}[H]
    \centering
    \includegraphics[width=0.8\textwidth]{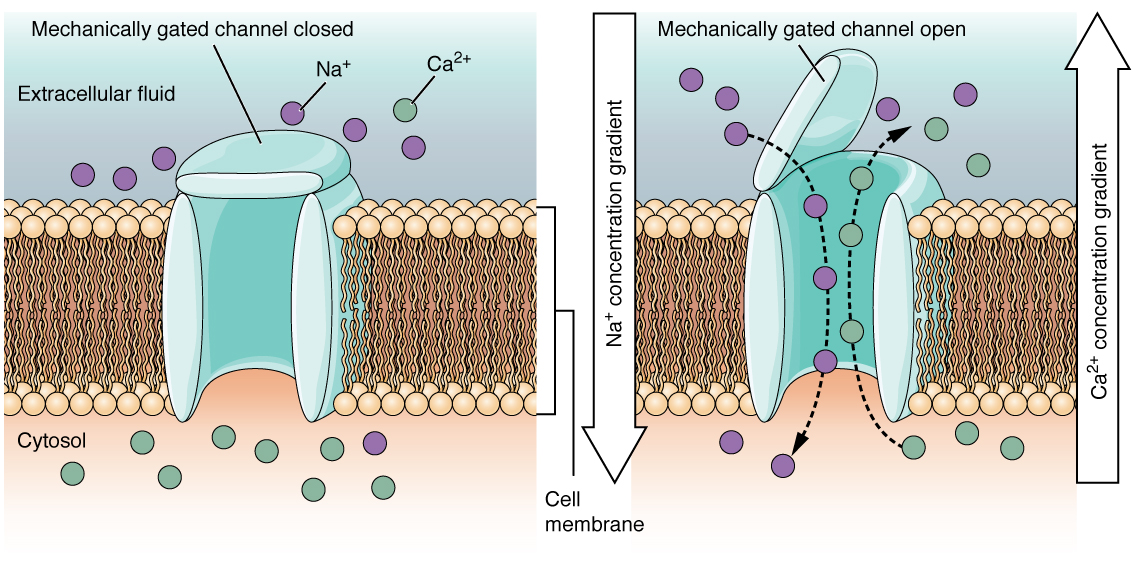}
    \caption{An Illustration of the Mechanically Gated Channel \cite{ion_gated_channel}.}
    \label{fig:mechanically_gated_channel-gated_channel}
\end{figure}

\end{itemize}

\subsubsection{Ion Pump}

Ion pumps are integral membrane proteins that carry out active transport, i.e., use cellular energy (ATP) to ``pump'' the ions against their concentration gradient. Such ion pumps take in ions from one side of the membrane (decreasing its concentration there) and release them on the other side (increasing its concentration there).

\begin{figure}[t]
    \centering
    \includegraphics[width=1.0\textwidth]{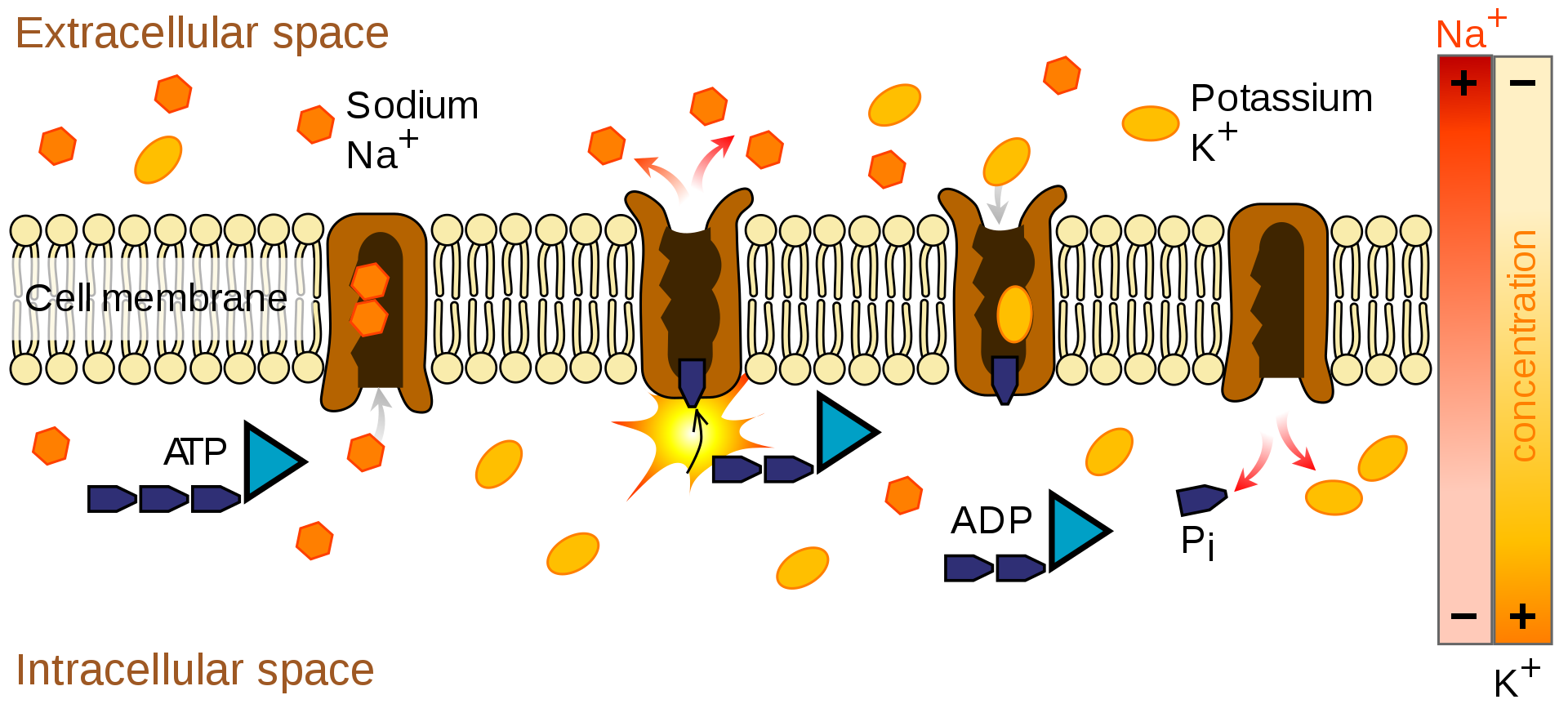}
    \caption{An Illustration of the Sodium-Potassium Pump \cite{membrane_potential} (The sodium-potassium pump uses energy derived from ATP to exchange sodium for potassium ions across the membrane).}
    \label{fig:sodium_potassium_pump}
\end{figure}

The ion pump most relevant to the action potential is the sodium-potassium pump as illustrated in Figure~\ref{fig:sodium_potassium_pump}, which transports three sodium ions out of the cell and two potassium ions in. As a consequence, the concentration of potassium ions K$^+$ inside the neuron is roughly 20-fold larger than the outside concentration, whereas the sodium concentration outside is roughly 9-fold larger than inside. In a similar manner, other ions have different concentrations inside and outside the neuron, such as calcium, chloride and magnesium.

If the numbers of each type of ion were equal, the sodium-potassium pump would be electrically neutral, but, because of the three-for-two exchange, it gives a net movement of one positive charge from intracellular to extracellular for each cycle, thereby contributing to a positive voltage difference. The pump has three effects: 
\begin{itemize}
\item It makes the sodium concentration high in the extracellular space and low in the intracellular space; 
\item It makes the potassium concentration high in the intracellular space and low in the extracellular space; 
\item It gives the intracellular space a negative voltage with respect to the extracellular space.
\end{itemize}

The sodium-potassium pump is relatively slow in operation. If a cell were initialized with equal concentrations of sodium and potassium everywhere, it would take hours for the pump to establish equilibrium. The pump operates constantly, but becomes progressively less efficient as the concentrations of sodium and potassium available for pumping are reduced.

Ion pumps influence the action potential only by establishing the relative ratio of intracellular and extracellular ion concentrations. The action potential involves mainly the opening and closing of ion channels not ion pumps. If the ion pumps are turned off by removing their energy source, or by adding an inhibitor such as ouabain, the axon can still fire hundreds of thousands of action potentials before their amplitudes begin to decay significantly. In particular, ion pumps play no significant role in the repolarization of the membrane after an action potential.

Another functionally important ion pump is the sodium-calcium exchanger. This pump operates in a conceptually similar way to the sodium-potassium pump, except that in each cycle it exchanges three Na$^+$ from the extracellular space for one Ca$^{2+}$ from the intracellular space. Because the net flow of charge is inward, this pump runs ``downhill'', in effect, and therefore does not require any energy source except the membrane voltage. Its most important effect is to pump calcium outward-it also allows an inward flow of sodium, thereby counteracting the sodium-potassium pump, but, because overall sodium and potassium concentrations are much higher than calcium concentrations, this effect is relatively unimportant. The net result of the sodium-calcium exchanger is that in the resting state, intracellular calcium concentrations become very low.

\subsection{Resting Potential, Graded Potential and Action Potential}

\begin{figure}[t]
    \centering
    \includegraphics[width=0.4\textwidth]{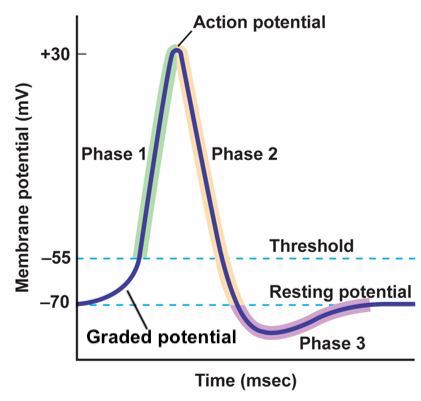}
    \caption{An Illustration of the Action Potential.}
    \label{fig:membrane_action_potential}
\end{figure}

In Figure~\ref{fig:membrane_action_potential}, we show the whole process for a neuron to file an impulse, which several different phases: resting potential, graded potential and action potential. Neuron membrane potential changes are associated with each of the phases. In this part, we will talk about the neuron membrane resting potential, graded potential and action potential, and introduce the working mechanism of the neurons

\subsubsection{Resting Potential}

According to \cite{resting_potential}, the relatively static membrane potential of quiescent cells is called the resting membrane potential (or resting voltage), as opposed to the specific dynamic electrochemical phenomena called action potential and graded membrane potential.

Apart from the latter two, which occur in excitable cells (neurons, muscles, and some secretory cells in glands), membrane voltage in the majority of non-excitable cells can also undergo changes in response to environmental or intracellular stimuli. In principle, there is no difference between resting membrane potential and dynamic voltage changes like action potential from a biophysical point of view: all these phenomena are caused by specific changes in membrane permeabilities for potassium, sodium, calcium, and chloride ions, which in turn result from concerted changes in functional activity of various ion channels, ion transporters, and exchangers. Conventionally, resting membrane potential can be defined as a relatively stable, ground value of transmembrane voltage in animal and plant cells.

In the case of the resting membrane potential across an animal cell's plasma membrane, potassium (and sodium) gradients are established by the Na$^+$/K$^+$-ATPase (sodium-potassium pump) which transports 2 potassium ions inside and 3 sodium ions outside at the cost of 1 ATP molecule. In other cases, for example, a membrane potential may be established by acidification of the inside of a membranous compartment (such as the proton pump that generates membrane potential across synaptic vesicle membranes).

In most neurons the resting potential has a value of approximately -70 mV. The resting potential is mostly determined by the concentrations of the ions in the fluids on both sides of the cell membrane and the ion transport proteins that are in the cell membrane. How the concentrations of ions and the membrane transport proteins influence the value of the resting potential is outlined below. The resting potential of a cell can be most thoroughly understood by thinking of it in terms of equilibrium potentials.

\subsubsection{Graded Potential}

As introduced in \cite{graded_potential}, graded potentials are changes in membrane potential that vary in size, as opposed to being all-or-none. They include diverse potentials such as receptor potentials, electrotonic potentials, subthreshold membrane potential oscillations, slow-wave potential, pacemaker potentials, and synaptic potentials, which scale with the magnitude of the stimulus. They arise from the summation of the individual actions of ligand-gated ion channel proteins, and decrease over time and space. They do not typically involve voltage-gated sodium and potassium channels. These impulses are incremental and may be excitatory or inhibitory. They occur at the postsynaptic dendrite in response to presynaptic neuron firing and release of neurotransmitter, or may occur in skeletal, smooth, or cardiac muscle in response to nerve input. The magnitude of a graded potential is determined by the strength of the stimulus.
\begin{itemize}
\item \textbf{Excitatory Postsynaptic Potentials}: Graded potentials that make the membrane potential less negative or more positive, thus making the postsynaptic cell more likely to have an action potential, are called excitatory postsynaptic potentials (EPSPs). Depolarizing local potentials sum together, and if the voltage reaches the threshold potential, an action potential occurs in that cell.

EPSPs are caused by the influx of Na$^+$ or Ca$^{2+}$ from the extracellular space into the neuron or muscle cell. When the presynaptic neuron has an action potential, Ca$^{2+}$ enters the axon terminal via voltage-dependent calcium channels and causes exocytosis of synaptic vesicles, causing neurotransmitter to be released. The transmitter diffuses across the synaptic cleft and activates ligand-gated ion channels that mediate the EPSP. The amplitude of the EPSP is directly proportional to the number of synaptic vesicles that were released.

If the EPSP is not large enough to trigger an action potential, the membrane subsequently repolarizes to its resting membrane potential. This shows the temporary and reversible nature of graded potentials.

\item \textbf{Inhibitory Postsynaptic Potentials}: Graded potentials that make the membrane potential more negative, and make the postsynaptic cell less likely to have an action potential, are called inhibitory post synaptic potentials (IPSPs). Hyperpolarization of membranes is caused by influx of Cl$^-$ or efflux of K$^+$. As with EPSPs, the amplitude of the IPSP is directly proportional to the number of synaptic vesicles that were released.

\end{itemize}

The resting membrane potential is usually around -70mV. The typical neuron has a threshold potential ranging from -40mV to -55mV. Temporal summation occurs when graded potentials within the postsynaptic cell occur so rapidly that they build on each other before the previous ones fade. Spatial summation occurs when postsynaptic potentials from adjacent synapses on the cell occur simultaneously and add together. An action potential occurs when the summated EPSPs, minus the summated IPSPs, in an area of membrane reach the cell's threshold potential.

\subsubsection{Action Potential}

As introduced in \cite{action_potential}, in physiology, an action potential occurs when the membrane potential of a specific axon location rapidly rises and falls: this depolarisation then causes adjacent locations to similarly depolarise. Action potentials occur in several types of animal cells, called excitable cells, which include neurons, muscle cells, endocrine cells, glomus cells, and in some plant cells.

In neurons, action potentials play a central role in cell-to-cell communication by providing for-or with regard to saltatory conduction, assisting-the propagation of signals along the neuron's axon toward synaptic boutons situated at the ends of an axon; these signals can then connect with other neurons at synapses, or to motor cells or glands. Action potentials in neurons are also known as ``nerve impulses'' or ``spikes'', and the temporal sequence of action potentials generated by a neuron is called its ``spike train''. A neuron that emits an action potential, or nerve impulse, is often said to ``fire''.

Action potentials are generated by special types of voltage-gated ion channels embedded in a cell's plasma membrane. These channels are shut when the membrane potential is near the (negative) resting potential of the cell which is usually -70mV for neurons, but they rapidly begin to open if the membrane potential increases to a precisely defined threshold voltage which is usually -55mV, depolarising the transmembrane potential. 

When the channels open, they allow an inward flow of sodium ions, which changes the electrochemical gradient, which in turn produces a further rise in the membrane potential. This then causes more channels to open, producing a greater electric current across the cell membrane and so on. The process proceeds explosively until all of the available ion channels are open, resulting in a large upswing in the membrane potential. The rapid influx of sodium ions causes the polarity of the plasma membrane to reverse, and the ion channels then rapidly inactivate. As the sodium channels close, sodium ions can no longer enter the neuron, and they are then actively transported back out of the plasma membrane. Potassium channels are then activated, and there is an outward current of potassium ions, returning the electrochemical gradient to the resting state. After an action potential has occurred, there is a transient negative shift, called the afterhyperpolarization.

In animal cells, there are two primary types of action potentials. One type is generated by voltage-gated sodium channels, the other by voltage-gated calcium channels. Sodium-based action potentials usually last for under one millisecond, but calcium-based action potentials may last for 100 milliseconds or longer. In some types of neurons, slow calcium spikes provide the driving force for a long burst of rapidly emitted sodium spikes. In cardiac muscle cells, on the other hand, an initial fast sodium spike provides a ``primer'' to provoke the rapid onset of a calcium spike, which then produces muscle contraction.

\subsection{Action Potential Phases and Ion Movement}

\begin{figure}[t]
    \centering
    \includegraphics[width=0.8\textwidth]{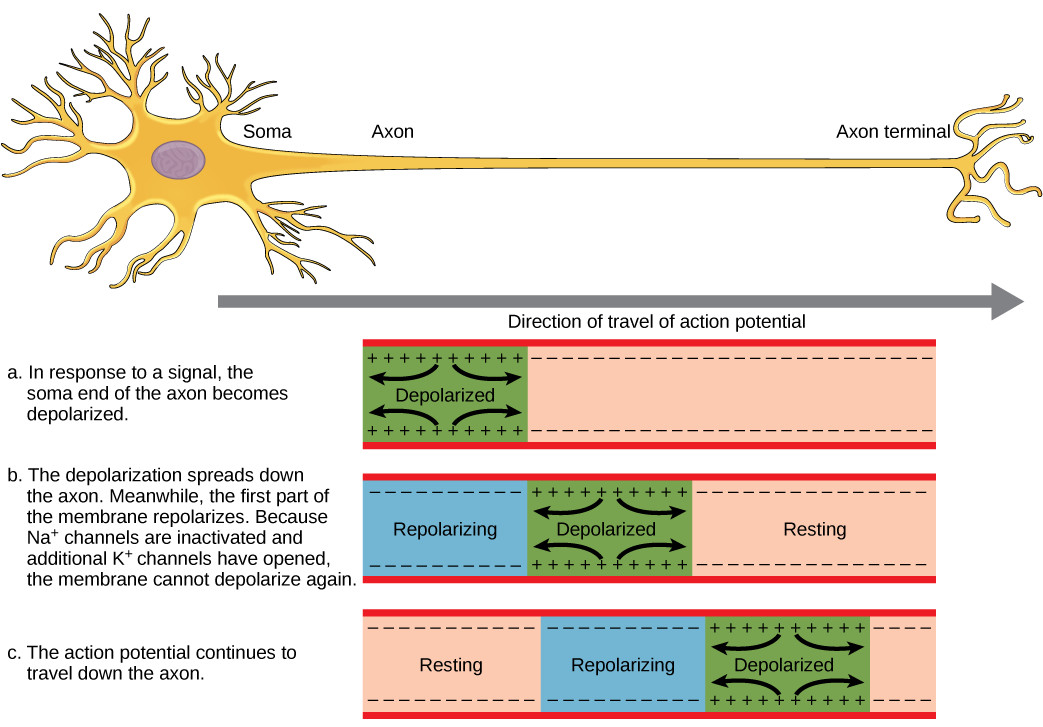}
    \caption{An Impulse Transmission via Potential Changes.}
    \label{fig:impulse_transmission}
\end{figure}

The course of the action potential can be divided into five parts: the rising phase, the peak phase, the falling phase, the undershoot phase, and the refractory period. During the rising phase the membrane potential depolarizes (becomes more positive). The point at which depolarization stops is called the peak phase. At this stage, the membrane potential reaches a maximum. Subsequent to this, there is a falling phase. During this stage the membrane potential becomes more negative, returning towards resting potential. The undershoot, or afterhyperpolarization, phase is the period during which the membrane potential temporarily becomes more negatively charged than when at rest (hyperpolarized). Finally, the time during which a subsequent action potential is impossible or difficult to fire is called the refractory period, which may overlap with the other phases. 

As illustrated in Figure~\ref{fig:impulse_transmission}, via such potential changes of different fractions of the axon, the impulse will be transmitted from one side to the other side along the axon. In addition, the myelin sheath cells on the axon further speed up the transmission of the impulse. To understand such a complex process, the readers are also suggested to refer to an online video offered by Harvard via \cite{action_potential_video} for more information.

\subsubsection{Stimulation and rising phase}

\begin{figure}[t]
    \centering
    \includegraphics[width=1.0\textwidth]{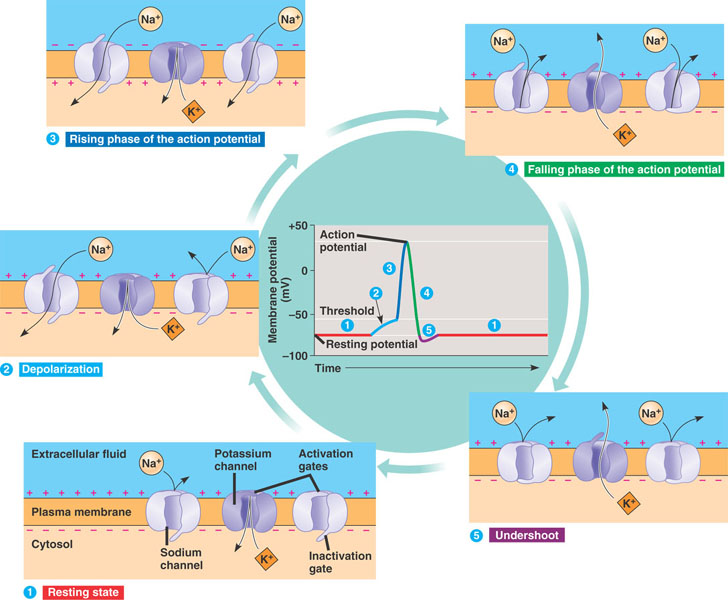}
    \caption{An Illustration of Ion Movement during an Action Potential. \cite{action_potential_diagram}.}
    \label{fig:membrane_potential_diagram}
\end{figure}

As illustrated in Figure~\ref{fig:membrane_potential_diagram}, a typical action potential begins at the axon hillock with a sufficiently strong depolarization, e.g., a stimulus that increases the membrane voltage. This depolarization is often caused by the injection of extra sodium cations into the cell; these cations can come from a wide variety of sources, such as chemical synapses, sensory neurons or pacemaker potentials.

For a neuron at rest, there is a high concentration of sodium (Na$^+$) and chloride (Cl$^-$) ions in the extracellular fluid compared to the intracellular fluid, while there is a high concentration of potassium (K$^+$) ions in the intracellular fluid compared to the extracellular fluid. The difference in concentrations, which causes ions to move from a high to a low concentration, and electrostatic effects (attraction of opposite charges) are responsible for the movement of ions in and out of the neuron. The inside of a neuron has a negative charge, relative to the cell exterior, from the movement of K$^+$ out of the cell. The neuron membrane is more permeable to K$^+$ than to other ions, allowing this ion to selectively move out of the cell, down its concentration gradient. This concentration gradient along with potassium leak channels present on the membrane of the neuron causes an efflux of potassium ions making the resting potential close to -70 mV. 

Since Na$^+$ ions are in higher concentrations outside of the cell, the concentration and voltage differences both drive them into the cell when Na$^+$ channels open. Depolarization opens both the sodium and potassium channels in the membrane, allowing the ions to flow into and out of the axon, respectively. If the depolarization is small (say, increasing the membrane voltage from -70 mV to -60 mV), the outward potassium current overwhelms the inward sodium current and the membrane repolarizes back to its normal resting potential around -70 mV. However, if the depolarization is large enough, the inward sodium current increases more than the outward potassium current and a runaway condition (positive feedback) results: the more inward current there is, the more the membrane voltage increases, which in turn further increases the inward current. 

\subsubsection{Peak and falling phase} 

According to Figure~\ref{fig:membrane_potential_diagram}, a sufficiently strong depolarization (increase in the membrane voltage to -55mV) causes the voltage-sensitive sodium channels to open; the increasing permeability to sodium drives the membrane voltage closer to the sodium equilibrium voltage. The increasing voltage in turn causes even more sodium channels to open, which pushes the membrane voltage still further upward. This positive feedback continues until the sodium channels are fully open. The sharp rise in the membrane voltage and sodium permeability correspond to the rising phase of the action potential.

The positive feedback of the rising phase slows and comes to a halt as the sodium ion channels become maximally open. At the peak of the action potential, the sodium permeability is maximized and the membrane voltage the membrane voltage is nearly equal to the sodium equilibrium voltage. However, the same raised voltage that opened the sodium channels initially also slowly shuts them off, by closing their pores; the sodium channels become inactivated. This lowers the membrane's permeability to sodium relative to potassium, driving the membrane voltage back towards the resting value. At the same time, the raised voltage opens voltage-sensitive potassium channels; the increase in the membrane's potassium permeability drives the membrane voltage down to -70mV. Combined, these changes in sodium and potassium permeability cause the membrane voltage to drop quickly, repolarizing the membrane and producing the ``falling phase'' of the action potential.

\subsubsection{Afterhyperpolarization}

As illustrated in Figure~\ref{fig:membrane_potential_diagram}, the depolarized voltage opens additional voltage-dependent potassium channels, and some of these do not close right away when the membrane returns to its normal resting voltage. In addition, further potassium channels open in response to the influx of calcium ions during the action potential. The intracellular concentration of potassium ions is transiently unusually low, making the membrane voltage the membrane voltage even closer to the potassium equilibrium voltage -70mV. The membrane potential goes below the resting membrane potential. Hence, there is an undershoot or hyperpolarization, termed an afterhyperpolarization, that persists until the membrane potassium permeability returns to its usual value, restoring the membrane potential to the resting state.

\subsubsection{Refractory period}

A cell that has just fired an action potential cannot fire another one immediately, since the Na$^+$ channels have not recovered from the deactivated state. The period during which no new action potential can be fired is called the absolute refractory period. At longer times, after some but not all of the ion channels have recovered, the axon can be stimulated to produce another action potential, but with a higher threshold, requiring a much stronger depolarization, e.g., to -30 mV. The period during which action potentials are unusually difficult to evoke is called the relative refractory period.

The absolute refractory period is largely responsible for the unidirectional propagation of action potentials along axons. At any given moment, the patch of axon behind the actively spiking part is refractory, but the patch in front, not having been activated recently, is capable of being stimulated by the depolarization from the action potential.

\section{Summary}

In this paper, we have introduce the basic neural units of the brain, including neurons, synapses and action potential. These basic units help illustrate the cell-level structure of the brain, as well has how impulses are propagated among neurons in the brain. In the next follow-up article \cite{zhang2019cognitive}, we will introduce the high-level cognitive functions (e.g., perception, attention and memory) of the brain.

\newpage

\vskip 0.2in
\bibliographystyle{plain}
\bibliography{reference}

\end{document}